\documentclass[aps,prb,twocolumn,floatfix]{revtex4}
\usepackage{graphicx, colordvi}
\usepackage{latexsym}
\usepackage{amsmath}
\usepackage{graphics}
\usepackage{amssymb}
\usepackage{layout}
\usepackage{verbatim}
\usepackage{amsfonts,epsfig}
\newcommand{\ket}[1]{\left|#1\right>}
\newcommand{\bra}[1]{\left< #1 \right|}
\begin{document}

\title{Kinetics of quasiparticle trapping in a Cooper-pair box}

\author{R. M. Lutchyn and L. I. Glazman}

\affiliation{ W.I.\ Fine Theoretical Physics Institute, University
of Minnesota, Minneapolis, Minnesota 55455, USA}

\date{\today }

\begin{abstract}
  We study the kinetics of the quasiparticle capture and emission
  process in a small superconducting island (Cooper-pair box)
  connected by a tunnel junction to a massive superconducting lead.
  At low temperatures, the charge on the box fluctuates
  between two states, even and odd in the number of
  electrons.  Assuming that the odd-electron state has
  the lowest energy, we evaluate the distribution of lifetimes of the
  even- and odd-electron states of the Cooper-pair box. The
  lifetime in the even-electron state is an exponentially
  distributed random variable corresponding to a homogenous
  Poisson process of ``poisoning'' the island with a
  quasiparticle. The distribution of lifetimes of the odd-electron
  state may deviate from the exponential one. The deviations come
  from two sources - the peculiarity
  of the quasiparticle density of states in a superconductor and the
  possibility of quasiparticle energy relaxation via phonon emission.
  In addition to the lifetime distribution, we also find spectral density of charge fluctuations generated by
  capture and emission processes. The complex statistics of the
  quasiparticle dwell times in the Cooper-pair box may result in strong deviations
  of the noise spectrum from the Lorentzian form.
\end{abstract}
\pacs{74.50.+r, 03.67.Lx, 03.65.Yz, 85.25.Cp}

\maketitle
\section{Introduction}
Properties of a mesoscopic superconducting circuit may depend
crucially on the presence of quasiparticles in its elements. The
operation of a superconducting charge qubit, for example, requires
two-electron periodicity of its charge states~\cite{Mannik,
Aumentado, Guillaume, Schneiderman, Turek, Naaman, Ferguson}. This
periodicity may be interrupted by the entrance of an unpaired
electron into the Cooper-pair box (CPB) serving as an active
element of a qubit. The quasiparticle changes the charge state of
CPB from even to odd, and lowers the charging energy. This
trapping phenomenon, commonly referred to as ``quasiparticle
poisoning'', is well-known from the studies of the charge parity
effect in superconductors\cite{Eiles, Matveev}. Quasiparticle
poisoning contributes to the phase relaxation in superconducting
qubits~\cite{Lutchyn2}. For a typical CPB size and tunnel
conductances of the order of unit quantum, the quasiparticle
dwelling times are of the order of a few $\mu$s. This time scale
is at the edge of accessibility for the modern
experiments~\cite{Lu}. Individual quasiparticle tunneling events
were resolved and the statistics of quasiparticle entrances and
exits from the CPB box were investigated in
Refs.~[\onlinecite{Naaman, Ferguson}]. The observed statistics of
entrances was well described by a standard Poissonian
process~\cite{Naaman,Ferguson}. For the quasiparticle exits, the
results are less clear. In many cases, it may be well described by
the Poissonian statistics~\cite{Naaman, Ferguson}. However, there
are indications of deviation from that simple law for some
samples~\cite{naaman-private-comm}.

In this paper, we develop a kinetic theory of quasiparticle
poisoning. We find the distribution of times $N_{\rm odd}(t)$ and
$N_{\rm even}(t)$ the CPB dwells, respectively, in odd- and
even-electron states. We also find the spectrum of charge noise
produced by the poisoning processes.

The conventional Poissonian statistics of the quasiparticle exits
would yield an exponential distribution for odd-electron lifetime
in the box. We see two reasons for the distribution function
$N_{\rm odd}(t)$ to deviate from that simple form. The first one
is related to the thermalization of a quasiparticle within the
CPB. If the rates of energy relaxation and of tunneling out for a
quasiparticle in CPB are of the same order, then two different
time scales control the short-time and long-time parts of the
distribution function $N_{\rm odd}(t)$. The shorter time scale is
defined by the escape rate $\Gamma_{\rm out}$ of the
unequilibrated quasiparticle from the CPB. The longer time scale
is defined by the rate of activation of equilibrated quasiparticle
to an energy level allowing an escape from CPB. The second reason
for the deviations from the exponential distribution controlled by
a single rate, comes from the singular energy dependence of the
quasiparticle density of states in a superconductor. Because of
it, the tunneling-out rate depends strongly on the quasiparticle
energy. Thus, even in the absence of thermalization the
quasiparticle escapes from CPB cannot be described by an
exponential distribution.

The conventional Poissonian statistics for both entrances and
exits of the quasiparticle would lead to a Lorentzian spectral
density $S_Q(\omega)$ of CPB charge fluctuations~\cite{Machlup}.
The interplay of tunneling and relaxation rates may result in
deviations from the Lorentzian function.  In the case of slow
quasiparticle thermalization rate compared to the quasiparticle
tunneling-out rate $\Gamma_{\rm out}$, the function $S_Q(\omega)$
roughly can be viewed as a superposition of two Lorentzians. The
width of the narrower one is controlled by the processes involving
quasiparticle thermalization and activation by phonons, while the
width of the broader one is of the order of the escape rate
$\Gamma_{\rm out}$.
\begin{figure}
\centering
\includegraphics[width=2.8in]{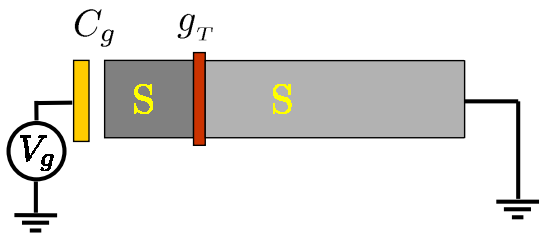}
\caption{(color online). Schematic picture of the Cooper-pair box
qubit. The left-hand side superconducting mesoscopic island is the
Cooper-pair box connected via a tunable Josephson junction to the
large superconducting lead (right-hand side). Gate bias is applied
through the capacitance $C_g$. The junction is characterized by
the dimensionless conductance $g_{_T}$.}\label{circuit}
\end{figure}

The paper is organized as follows. We begin in
Sec.~\ref{theomodel} with the qualitative derivation and
discussion of the main results. In the next sections
(\ref{even}-\ref{odd}) we derive and solve the microscopic master
equations for the kinetics of the quasiparticle capture and
emission, and calculate the lifetime distribution functions in the
even- and odd-charge states. In Sec.~\ref{secnoise} we calculate
charge noise spectral density $S_Q(\omega)$ for the Cooper-pair
box. In Sec.~\ref{conclusion} we summarize the main results. Some
technical details are relegated to the Appendix.

\section{Qualitative considerations and main results.}\label{theomodel}
\subsection{Relevant time scales.}

Dynamics of the Cooper-pair box coupled to the superconducting
lead through the Josephson junction, see Fig.~\ref{circuit}, is
described by the Hamiltonian
\begin{equation}\label{Hqubit}
H=H_{_C}+H^b_{\rm{BCS}}+H^l_{\rm{BCS}}+H_{_T}.
\end{equation}
Here $H^b_{\rm{BCS}}$ and $H^l_{\rm{BCS}}$ are BCS Hamiltonians
for box and the lead; $H_{_C}=E_c(\hat{Q}/e\!-\!N_g)^2$, with
$E_c$, $N_g$, and $\hat{Q}$ being the charging energy,
dimensionless gate voltage, and charge of the CPB, respectively.
The tunneling Hamiltonian $H_{_T}$ is defined in the conventional
way
\begin{equation}\label{HT}
H_{_T}=\sum_{kp\sigma}\left(t_{kp}c_{k,\sigma}^{\dag}c_{p,\sigma}+\emph{\rm{H.c.}}\right),
\end{equation}
where $t_{kp}$ is the tunneling matrix element, $c_{k,\sigma}$ and
$c_{p,\sigma}$ are the annihilation operators for an electron in
the state $\ket{k,\sigma}$ in the CPB and state $\ket{p,\sigma}$
in the superconducting lead, respectively. Here superconducting
gap energy is the largest energy scale, $\Delta> E_c> E_{_J} \gg
T$. In order to distinguish between Cooper pair and quasiparticle
tunneling, we present the Hamiltonian~(\ref{Hqubit}) in the
form~\cite{Lutchyn1}
\begin{equation}\label{Hqubit2}
H=H_0+V,\mbox{ and }V=H_{_T}\!-\!H_{_{_J}}.
\end{equation}
Here $H_0=H_{_C}+H^b_{\rm{BCS}}+H^l_{\rm{BCS}}+H_{_J}$, and
$H_{_J}$ is the Hamiltonian describing Josephson tunneling
\begin{equation}
H_{_J}=\ket{N}\bra{N} H_{_T} \frac{1}{E-H_0}H_{_T}
\ket{N\!+2}\bra{N\!+2} +{\rm H.c.}\nonumber
\end{equation}
The matrix element $\bra{N} H_{_T} \frac{1}{E-H_0}H_{_T}
\ket{N+2}$ is proportional to the Josephson energy $E_{_J}$. (Here
$E_{_J}$ is given by the Ambegaokar-Baratoff relation.) The
perturbation Hamiltonian $V$ defined in Eq.~(\ref{Hqubit2}) is
suitable for calculation of the quasiparticle tunneling rate.
\begin{figure}
\centering
\includegraphics[width=2.6in]{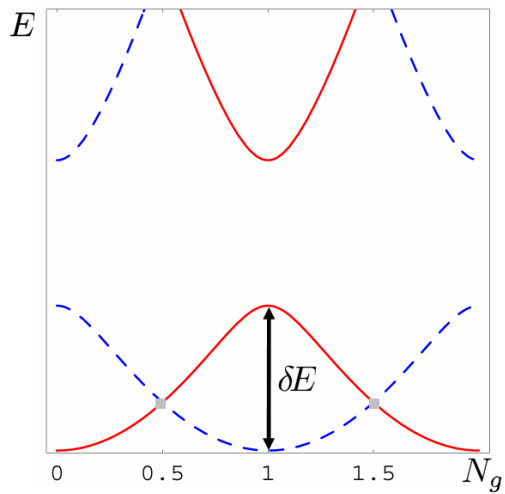}
\caption{(color online). Energy of the Cooper-pair box as a
  function of dimensionless gate voltage $N_g$ in units of $e$.
  Solid (red) line corresponds to even-charge state of the box, dashed (blue) line
  corresponds to the odd-charge state of the box. The trap depth $\delta
  E$ is the ground state energy difference between the even-charge state (no
quasiparticles in the CPB), and odd-charge state (an unpaired
electron in the CPB) at $N_g=1$. (We assume here equal gap
energies in the box and the lead,
$\Delta_{l}=\Delta_{b}=\Delta$.)} \label{fig1}
 \end{figure}

Energy of the system as a function of the gate voltage is shown in
Fig.~\ref{fig1}. At $N_g=1$ the electrostatic energy of the system
is minimized when unpaired electron resides in the CPB. Thus, at
$N_g=1$ the CPB is a trap for a quasiparticle. The trap depth
$\delta E$ is equal the ground state energy difference between the
even-charge state (no quasiparticles in the CPB) and odd-charge
state (an unpaired electron in the CPB). For equal gap energies in
the box and the lead, $\Delta_{l}=\Delta_{b}=\Delta$, the trap is
formed due to the Coulomb blockade effect. In the case $E_c\gg
E_{_J}/2$ one has
\begin{equation}\label{depth}
\delta E \approx E_c-\frac{E_{_J}}{2}\gg T,
\end{equation}
and only two lowest charge states are important, see
Fig.~\ref{fig1}. Also, we assume that there is at most one
quasiparticle in the box in the odd state~\cite{footnote}.

The transition probability per unit time between odd and
even-charge states $W(p,k)$ can be obtained using the Fermi golden
rule ($\hbar=1$),
\begin{eqnarray}\label{rates}
W(p,k)\!=\!2\pi|\bra{p,{\rm e}}V\ket{{\rm
o},k}|^2\delta(E_p\!+\!\delta E\!-\!E_k).
\end{eqnarray}
Here the state $\ket{{\rm e},p}$ corresponds to even-charge state
of the box and the quasiparticle in the state $\ket{p\,}$ in the
reservoir; the state $\ket{{\rm o},k}$ corresponds to the
odd-charge state of the box and quasiparticle in the state
$\ket{k\,}$ within the box. The quasiparticle energies in the CPB
and lead $E_{k/p}$ are defined as
$E_{k/p}=\sqrt{\xi_{k/p}^2+\Delta^2}$. Matrix elements
$\bra{p,{\rm e}}V\ket{{\rm o },k}$ can be calculated using the
Bogoliubov transformation~\cite{Schrieffer, Tinkham}. Taking into
account the relation between tunneling matrix elements and the
normal-state junction conductance the expression for $W(p,k)$ can
be written as
\begin{eqnarray}\label{rates2}
W(p,k)\!=\!\frac{g_{_{\rm{T}}}\delta_l
\delta_b}{8\pi}\left(\!1\!+\!\frac{\xi_p\xi_k\!-\!\Delta^2}{E_pE_k}\!\right)\!\delta(E_p\!+\!\delta
E\!-\!E_k)
\end{eqnarray}
with $\delta_{b/l}$ being mean level spacing in the box/lead, and
$g_{_{\rm{T}}}$ being dimensionless conductance of the junction.
\begin{figure}
\centering
\includegraphics[height=2.5in]{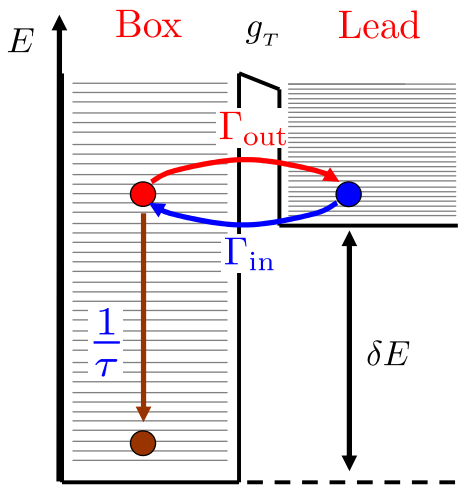}
\caption{(color online). Schematic picture of the CPB-lead system
showing allowed transitions for the quasiparticle injected into
the excited state of the box. At $N_g\!=\!1$ the Cooper-pair box
is a trap for quasiparticle.} \label{fig2}
 \end{figure}

\begin{figure}
\centering
\includegraphics[width=3.3in, scale=1]{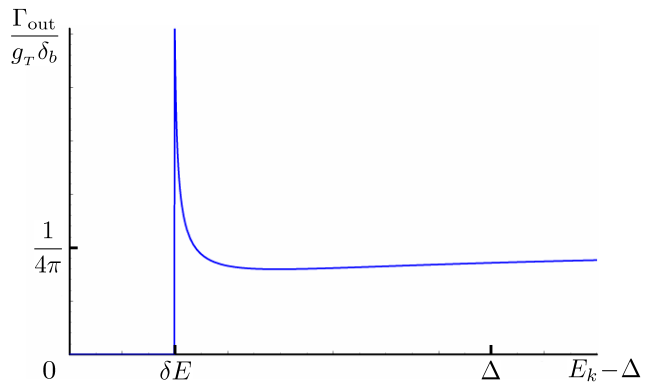}
\caption{The dependence of the escape rate
$\Gamma_{\rm{out}}(E_k)$ on energy $E_k$.} \label{fig_gammaout}
 \end{figure}

Using the transition rate~(\ref{rates2}), one can calculate the
level width of the state $\ket{{\rm o },k}$ with respect to
quasiparticle tunneling through the junction to the lead,
\begin{eqnarray}\label{gammaout}
\!\!\Gamma_{\rm{out}}(\!\! &\!\!E_k\!\!&\!\!) \equiv\sum_{p}W(p,k)\\
\!&\!=\!&\!\frac{g_{_{\rm{T}}}\delta_b}{4\pi}
\frac{(E_k\!-\!\delta E)E_k\!-\!\Delta^2}{(E_k\!-\!\delta
E)E_k}\nu(E_k\!-\!\delta
E)\Theta(\!E_k\!-\!E_{\rm{thd}}).\nonumber
\end{eqnarray}
The Heaviside function $\Theta(x)$ appears in Eq.~(\ref{gammaout})
because there are no states to tunnel into for a quasiparticle
with energy lower than the threshold energy $E_{\rm{thd}}$, see
Fig.~\ref{fig2},
\begin{equation}\label{threshold}
E_{\rm{thd}}=\Delta +\delta E.
\end{equation}
The quasiparticle density of states $\nu(E_k)$ (in units of the
normal density of states at the Fermi level) is given by
\begin{equation}\label{DOS}
\nu(E_k)=\frac{E_k}{\sqrt{E_k^2-\Delta^2}}.
\end{equation}
Due to the square-root singularity here, the rate
$\Gamma_{\rm{out}}(E_k)$ has square-root divergence at
$E_k=E_{\rm{thd}}$, see Fig.~(\ref{fig_gammaout}).

The quasiparticle may enter and subsequently leave the Cooper-pair
box without changing its energy. For such elastic process, the
excess energy of the exiting quasiparticle is equal to its initial
energy, and is of the order of the temperature, \emph{i.e.},
$E_k\!-\!E_{\rm{thd}}\!\sim\! T$. Therefore, the corresponding
escape rate is
\begin{equation}\label{Gammaout0}
\Gamma_{\rm{out}}=\frac{g_{_{\rm{T}}}\delta_b}{4\pi}\nu(T)\frac{\delta
E}{\delta E+\Delta}.
\end{equation}
Here for brevity we denote
$\nu(T)\equiv\nu(E_k\!=\!\Delta\!+\!T)$. For the system with
$g_{_{\rm{T}}}\lesssim 1$, volume of the CPB $V_b \lesssim 1\mu
m^3$, temperature $T \sim 50$mK and $\delta E\sim 0.5$K, the
typical escape time $\Gamma_{\rm{out}}^{-1}$ is of the order of a
$\mu s$.

To find the average rate $\Gamma_{\rm{in}}$ of quasiparticle
tunneling from the lead to the CPB, we integrate the transition
probability per unit time~(\ref{rates2}) with the distribution
function $f(E_p)$ of quasiparticles in the lead,
\begin{equation}\label{Gamma_in}
\Gamma_{\rm{in}}=\sum_{p,k} W(k,p)f(E_p).
\end{equation}

Upon elastically tunneling into the excited state in the CPB the
quasiparticle can relax to the bottom of the trap, see
Fig.~\ref{fig2}. For that, the quasiparticle needs to give away
energy $\sim \delta E$. At low temperatures the dominant mechanism
of quasiparticle energy relaxation is due to electron-phonon
inelastic scattering rate $1/\tau(E_k)$. At low temperature
quasiparticles are tunneling into the box through the energy
levels just above the threshold energy $E_k\sim E_{\rm{thd}}$, see
Eq.~(\ref{threshold}). Assuming $\delta E\ll \Delta$, the typical
quasiparticle relaxation time $\tau$ is given by~\cite{Kaplan}
\begin{eqnarray}\label{tau}
\tau\equiv\tau(E_k\!\sim\!
E_{\rm{thd}})\!\approx\tau_0\left(\frac{\Delta}{T_c}\right)^{-3}\left(\frac{\delta
E}{\Delta}\right)^{-\frac{7}{2}}.
\end{eqnarray}
Here $\tau_0$ is a characteristic parameter defining the average
electron-phonon scattering rate at $T=T_c$ with $T_c$ being
superconducting transition temperature. In aluminum, a typical
material used for CPB, $\tau_0\approx 0.1-0.5\,\mu
s$~\cite{Kaplan, Clarke, Santhanam}. As one can see from
Eq.~(\ref{tau}), the quasiparticle relaxation rate is a strong
function of the trap depth $\delta E$. Therefore, depending on
$\delta E$ there are two kinds of traps - ``shallow" traps
corresponding to $\tau \Gamma_{\rm{out}}\gg 1$, and ``deep" traps
with $\tau \Gamma_{\rm{out}}\ll 1$. (Note, for shallow traps we
still assume $\delta E \gg T$.) The important quantity
characterizing the traps is the probability $P_{\rm{tr}}$ for a
quasiparticle to relax to the bottom of the trap before an escape,
\begin{eqnarray}\label{Ptr}
P_{\rm{tr}}=\frac{1/\tau}{1/\tau+\Gamma_{\rm{out}}}.
\end{eqnarray}

\subsection{Lifetime distribution function.}

 Experimentally observable quantity~\cite{Naaman, Ferguson}, which reveals
the kinetics of quasiparticle trapping, is the lifetime
distribution function $N_{\rm{odd}}(t)$ of odd-charge states of
the CPB. The distribution of lifetimes $N_{\rm{odd}}(t)$ depends
on the internal dynamics of the quasiparticle in the CPB,
\emph{i.e} the ratio of $\tau\Gamma_{\rm{out}}$.

We start with the discussion of the long-time asymptote of the
lifetime distribution function. At $t>\tau$ the dwell-time
distribution $N_{\rm{odd}}(t)$ is governed by phonon-assisted
activation of the thermalized quasiparticle in the trap. The
phonon adsorption processes are statistically independent from
each other. Hence, the lifetime distribution exponentially decays
with time
\begin{eqnarray}\label{tail}
N_{\rm{odd}}(t)\propto \exp(-\gamma t)
\end{eqnarray}
with the rate
\begin{eqnarray}\label{qual_gamma}
\gamma\approx \frac{1}{\tau}\frac{\nu(\delta
E)}{\nu(T)}\exp\left(-\frac{\delta
E}{T}\right)\left(1-P_{\rm{tr}}\right).
\end{eqnarray}
This expression can be understood as follows. The rate of thermal
activation of the quasiparticle from the bottom of the trap to the
threshold energy is $\frac{1}{\tau}\frac{\nu(\delta
E)}{\nu(T)}\exp\left(-\frac{\delta E}{T}\right)$, for brevity we
define $\nu(\delta E)\!\equiv\!\nu(E_k\!=E_{\rm{thd}})$. The
additional factor $\nu(\delta E)/\nu(T)$ here comes from the
difference of the quasiparticle density of states at the bottom of
the trap $\nu(T)$ and at the threshold energy $\nu(\delta E)$. The
last term $\left(1-P_{\rm{tr}}\right)$ in Eq.~(\ref{qual_gamma})
corresponds to the probability of the quasiparticle escape to the
lead upon activation. Equation~(\ref{qual_gamma}) allows us to
consider limiting cases of $\tau\Gamma_{\rm{out}}\ll 1$ and
$\tau\Gamma_{\rm{out}}\gg 1$.

\begin{figure}
\centering
\includegraphics[height=3.4in]{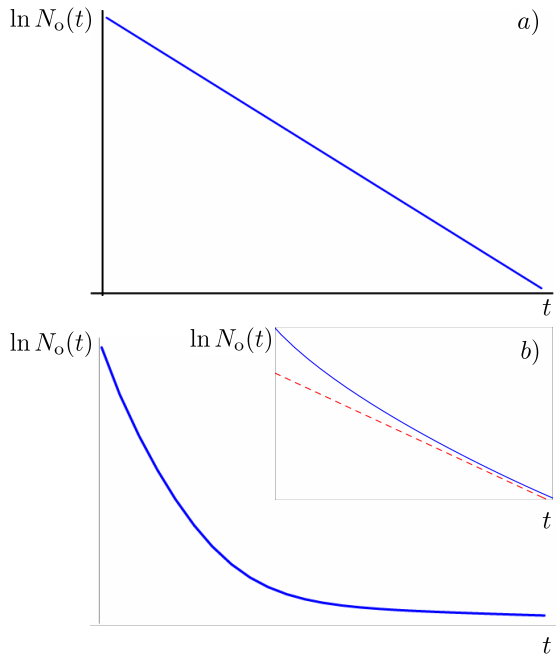}
\caption{(color online). a) Schematic picture of the lifetime
distribution function for ``deep" traps ($\tau\Gamma_{\rm{out}}\ll
1$). b) Schematic picture of the lifetime distribution function
for ``shallow" traps ($\tau\Gamma_{\rm{out}}\gg 1$). Inset:
Deviations of $N_{\rm{odd}}(t)$ from exponential distribution at
short times.}\label{lifetime}
\end{figure}

In the case of ``deep" traps ($\tau\Gamma_{\rm{out}}\ll 1$) most
quasiparticles upon entering the excited state in the box quickly
thermalize . Therefore, the main contribution to lifetime
distribution function comes from phonon-assisted escapes described
by Eq.~(\ref{tail}), see Fig.~(\ref{lifetime}). The activation
escape rate of Eq.~(\ref{qual_gamma}) in this limit equals
\begin{eqnarray}\label{qual_rate_fast}
\gamma_{\rm{f}}\approx\Gamma_{\rm{out}}\frac{\nu(\delta
E)}{\nu(T)}\exp\left(-\frac{\delta E}{T}\right),
\end{eqnarray}
since $1-P_{\rm{tr}}\simeq \Gamma_{\rm{out}}\tau$, see
Eq.~(\ref{Ptr}).

 In the opposite limit $\tau\Gamma_{\rm{out}}\gg 1$,
\emph{i.e.} ``shallow" traps, the probability for a quasiparticle
to relax to the bottom of the trap is small $P_{\rm{tr}}\ll 1$.
Therefore, upon elastically tunneling into the excited state in
the CPB the quasiparticles will predominantly return to the
reservoir unequilibrated. Nevertheless, there is a small fraction
of quasiparticles~($\sim1/\tau\Gamma_{\rm{out}}$) that do relax to
the bottom of the trap, and stay in the box much longer than
unequilibrated ones. Thus, at $t>\tau$ the dwell-time distribution
function $N_{\rm{odd}}(t)$ has an exponentially decaying
tail~(\ref{tail}), see Fig.~(\ref{lifetime}), with
phonon-activated escape rate
\begin{eqnarray}\label{qual_rate_slow}
\gamma_{\rm{s}}\approx\frac{1}{\tau}\frac{\nu(\delta
E)}{\nu(T)}\exp\left(-\frac{\delta E}{T}\right).
\end{eqnarray}
At $t\!\sim\!\tau$ the typical value of the lifetime distribution
function is $N_{\rm{odd}}(t\!\sim\!\tau)\sim
\gamma_{\rm{s}}/\tau\Gamma_{\rm{out}}$.

At short times, $t\ll\tau$, the lifetime distribution function
$N_{\rm{odd}}(t)$ describes the kinetics of unequilibrated
quasiparticles. Quasiparticles tunnel into the box through the
energy levels $E_k=E_{\rm{thd}}+\varepsilon$ (here
$\varepsilon\geq 0$), and predominantly reside there until the
escape with the rates $\Gamma_{\rm{out}}(\varepsilon)$. For a
given energy level $\varepsilon$ the lifetime distribution is
exponential
\begin{eqnarray}
N_{\rm{odd}}(\varepsilon, t)\propto
\exp(-\Gamma_{\rm{out}}(\varepsilon)t).
\end{eqnarray}
Note that upon entering into the CPB from the reservoir the
quasiparticles can populate different levels within the energy
strip $\sim T$, see Eq.~(\ref{IC}). Therefore, experimentally
observable quantity $N_{\rm{odd}}(t)$, obtained by the statistical
averaging over a large number of the tunneling events, is given by
\begin{eqnarray}
N_{\rm{odd}}(t)\propto\int_{0}^{\infty}d\varepsilon
\exp\left(-\frac{\varepsilon}{T}-\Gamma_{\rm{out}}(\varepsilon)t\right).
\end{eqnarray}
Taking into account the singularity of
$\Gamma_{\rm{out}}(\varepsilon)$ at small energies
$\Gamma_{\rm{out}}(\varepsilon)\propto \varepsilon^{-1/2}$, we
find that $N_{\rm{odd}}(t)$ deviates from the simple exponential
distribution [see Fig.~(\ref{lifetime})],
\begin{eqnarray}
N_{\rm{odd}}(t)\propto\exp\left(-3\left(\frac{\Gamma_{\rm{out}}t}{2}\right)^{2/3}\right)
\end{eqnarray}
at times $t \gtrsim 1/\Gamma_{\rm{out}}$. See also
Sec.~(\ref{odd}) for more details.

\subsection{Charge Noise Power Spectrum.}

Anomalies in the lifetime distribution, see Fig.~(\ref{lifetime}),
should also lead to a specific spectrum of charge fluctuations. We
define the spectral density of charge fluctuations $S_{Q}(\omega)$
in the Cooper-pair box as
\begin{eqnarray}\label{noise_def0}
S_{Q}(\omega)\!=\!\int_{-\infty}^{\infty}dte^{i\omega
t}\left[\langle \delta Q(t)\delta Q(0)\rangle\!+\!\langle \delta
Q(0)\delta Q(t)\rangle\right]\nonumber\\
\end{eqnarray}
with $\delta Q(t)=Q(t)-\langle Q\rangle$. The variance of the
fluctuations of charge $Q$ in the CPB,
\begin{eqnarray}\label{variance}
\langle \delta Q^2\rangle=\int_{0}^{\infty}\frac{d\omega}{2\pi}
S_{Q}(\omega),
\end{eqnarray}
is a thermodynamic, not a kinetic, quantity, and is known from
statistical mechanics. The kinetics of the system is reflected in
the dependence of the noise spectrum~(\ref{noise_def0}) on the
frequency $\omega$.

In the limit of fast relaxation $\tau\Gamma_{\rm{out}}\ll 1$ the
escapes from the CPB are given by one time scale
(\ref{qual_rate_fast}). The quasiparticle entrances into and exits
from the CPB are random, and can be described by Poisson
processes. Thus, $S_Q(\omega)$ is given by the Lorentzian function
corresponding to random telegraph noise~\cite{Machlup},
\begin{eqnarray}\label{qual_noise_fast}
S_Q(\omega)\approx
4e^2\bar{\sigma}_{\rm{odd}}(1-\bar{\sigma}_{\rm{odd}})\frac{\tau_{\rm{eff}}}{(\omega\tau_{\rm{eff}})^2+1}.
\end{eqnarray}
Here $\bar{\sigma}_{\rm{odd}}$ is an equilibrium average
occupation of the odd-charge state in the CPB
($0\leq\bar{\sigma}_{\rm{odd}}\leq 1$), see
Eq.~(\ref{stationary1}) for details. At low temperature ($T\ll
\delta E$) the box is predominantly in the odd-charge state,
\emph{i.e} $(1-\bar{\sigma}_{\rm{odd}})\propto \exp(-\delta E/T)$.
The rate of activated quasiparticle escape processes has the same
small exponent, therefore the width of the
Lorentzian~(\ref{qual_noise_fast}) is mainly given by the
transitions from even to odd-electron state,
\begin{eqnarray}\label{qual_width_fast}
\frac{1}{\tau_{\rm{eff}}}\approx\Gamma_{\rm{in}}.
\end{eqnarray}
See Eq.~(\ref{width_fast}) for the full result.

In the limit of slow relaxation ($\tau\Gamma_{\rm{out}}\gg 1$) the
charge noise power spectrum $S_Q(\omega)$ deviates significantly
from the Lorentzian. These deviations stem from the fact that a
quasiparticle may escape from the box before or after the
equilibration, which results in two characteristic time scales for
the escapes~\cite{Uren}, see Fig.~\ref{lifetime}. Consequently,
the function $S_Q(\omega)$ can be roughly viewed as a
superposition of two Lorentzians, and is similar to carrier
concentration fluctuations in semiconductors due to
trapping~\cite{Lax_Mengert}. The ``narrow" Lorentzian describes
the dynamics of slow fluctuations due to phonon-assisted trapping
of quasiparticles
\begin{eqnarray}\label{qual_noise_slow1}
S_Q^{(1)}(\omega)\sim
e^2\frac{\bar{\sigma}_{\rm{odd}}(1-\bar{\sigma}_{\rm{odd}})\tau_{\rm{eff}}}{(\omega\tau_{\rm{eff}})^2+1},\mbox{\,
}\frac{1}{\tau_{\rm{eff}}}\!\approx\!\frac{1}{\tau}\frac{\Gamma_{\rm{in}}}{\Gamma_{\rm{in}}\!+\!\Gamma_{\rm{out}}}.
\end{eqnarray}
The width $\tau_{\rm{eff}}^{-1}$ here is determined by the
probability of quasiparticle trapping per unit time. (Like above,
we assume here $T\ll \delta E$ and neglect activated escape rate.)
The second (quasi) Lorentzian function $S_Q^{(2)}(\omega)$ is
associated with fast charge fluctuations reflecting the kinetics
of unequilibrated quasiparticles. Assuming
$\omega\gg\Gamma_{\rm{out}}\gg\Gamma_{\rm{in}}$ the asymptote of
$S_Q^{(2)}(\omega)$ is
\begin{eqnarray}\label{qual_noise_slow2}
S_Q^{(2)}(\omega)\sim
e^2\frac{\bar{\sigma}_{\rm{odd}}}{\Gamma_{\rm{out}}}\exp(-\delta
E/T)\left(\frac{\Gamma_{\rm{out}}}{\omega}\right)^2.
\end{eqnarray}
The width of $S_Q^{(2)}(\omega)$ is determined by the typical
escape rate of unequilibrated quasiparticles from the box
$\Gamma_{\rm{out}}$ defined in Eq.~(\ref{Gammaout0}). Similar to
the lifetime distribution, see Fig.~\ref{lifetime}, we predict
deviations of $S_Q^{(2)}(\omega)$ from the Lorentzian function at
$\omega\!\sim\!\Gamma_{\rm{out}}$ due to the peculiarity of the
quasiparticle density of states.

The high-frequency tail of $S_Q(\omega)$ is provided by
Eq.~(\ref{qual_noise_slow2}). However, the contribution of
$S_Q^{(2)}(\omega)$ to the sum rule~(\ref{variance}) is much
smaller than that from $S_Q^{(1)}(\omega)$. In other words, the
main contribution to the noise power comes from slow fluctuations.
It resembles the case of the current noise in superconducting
detectors~\cite{Prober}.

In the rest of the paper, we provide detailed derivation of the
results discussed qualitatively in this section.

\section{Lifetime distribution of the even-charge state.}\label{even}

Let us assume that the system switched to the even state at $t=0$,
and introduce the probability density $N_{\rm{even}}(E_k,t)$ for a
quasiparticle to enter the CPB for the first time through the
state $E_k$. Then, the probability density for the CPB to reside
in the even state until time $t$ is
\begin{eqnarray}
N_{\rm{even}}(t)=\sum_k N_{\rm{even}}(E_k,t).
\end{eqnarray}
$N_{\rm{even}}(E_k,t)$ is given by the conditional probability of
quasiparticle entering the CPB into an empty state $E_k$ during
the interval $(t,t+dt)$ times the probability that any
quasiparticle has not entered into any state in the CPB during the
preceding interval $(0,t)$,
\begin{eqnarray}\label{qin}
N_{\rm{even}}(E_k,t)dt&=&\sum_p W(k,p)f(E_p)\nonumber\\
\!&\!\times\!&\!\left(1-\sum_{k'}
\int_0^td{t'}N_{\rm{even}}(E_{k'},t')\right)\!dt.
\end{eqnarray}
Summing Eq.~(\ref{qin}) over states $k$ and solving for
$N_{\rm{even}}(t)$ one finds
\begin{eqnarray}
N_{\rm{even}}(t)=\Gamma_{\rm{in}} \exp\left(-\Gamma_{\rm{in}}
t\right),
\end{eqnarray}
which corresponds to a homogenous Poisson process. The
quasiparticle tunneling rate from the lead to the CPB
$\Gamma_{\rm{in}}$ is given by Eq.~(\ref{Gamma_in}).

Recent experiments by Aumentado \emph{et.
al.}~[\onlinecite{Aumentado,Naaman}] indicate that the density of
quasiparticles $n^l_{\rm{qp}}$ in the lead exceeds the equilibrium
one at the temperature of the cryostat. The origin of
nonequilibrium quasiparticles is not clear, but it is plausible to
assume that quasiparticle distribution function in the lead
$f(E_p)$ is given by the Boltzman function
\begin{eqnarray}\label{distr_func}
f(E_p)=\exp\left(-\frac{E_p-\mu_l}{T}\right)
\end{eqnarray}
with some effective chemical potential and temperature, $\mu_l$
and $T$, respectively. The chemical potential $\mu_l$ is related
to the quasiparticle density by the equation
\begin{eqnarray}\label{density}
n^l_{\rm{qp}}=\frac{1}{V_l}\sum_{p}f(E_p).
\end{eqnarray}
Here $V_l$ is the volume of the lead. We consider the density of
quasiparticles $n^l_{\rm{qp}}$ and their effective temperature as
input parameters here, which can be estimated from the
experimental data~\cite{Aumentado, Naaman, Ferguson}. Taking into
account Eq.~(\ref{distr_func}) we can evaluate the right-hand side
of Eq.~(\ref{Gamma_in}) to obtain
\begin{eqnarray}\label{Gamma_in2}
\Gamma_{\rm{in}}=\frac{g_{_{\rm{T}}}n^l_{\rm{qp}}}{4\pi
\nu_{_F}}\nu(\delta E)\frac{\delta E}{\Delta+\delta E}.
\end{eqnarray}
Here $\nu_F$ is the normal density of states at the Fermi level.

The average waiting time in the even-charge state is
\begin{eqnarray}
\langle T_{\rm{even}} \rangle=\int_0^{\infty} N_{\rm{even}}(t)t
dt=\Gamma_{\rm{in}}^{-1}.
\end{eqnarray}
This result is expected for the conventional Poisson process.

\section{Lifetime distribution of the
odd-charge state.}\label{odd}

\subsection{Master equation for survival probability.}
The distribution of dwell times for odd-charge state is more
complicated than for even-charge state due to the internal
dynamics of the quasiparticle in the CPB. Upon tunneling
elastically into the box the quasiparticle enters into the excited
state with typical excess energy $\delta E$ above the gap in the
island. The dwell time of the quasiparticle in the box depends
whether upon tunneling into the excited state it relaxes to the
bottom of the trap or tunnels out un-equilibrated, see Fig. 2. In
order to describe the physics of quasiparticle tunneling we
develop a formalism similar to the rate equations theory. We will
include electron-phonon collision integral into our equations to
account for the internal dynamics of the quasiparticle inside the
CPB. The experimentally accessible quantity is the probability
density $N_{\rm{odd}}(t)$ of leaving an odd state in the time
interval ($t,t+dt$) assuming that the quasiparticle resided
continuously in the box during the time interval $(0,t)$. The
object convenient for evaluation is the conditional probability
$S_{\rm{odd}}(t)$ (or survival probability) for a quasiparticle to
occupy a given level, under the condition that the unpaired
electron continuously resided in the box over the time interval
$(0,t)$. The lifetime distribution $N_{\rm{odd}}(t)$ can be easily
obtained from $S_{\rm{odd}}(t)$,
\begin{eqnarray}\label{dwell_distribution}
N_{\rm{odd}}(t)=\frac{d}{dt}(1-S_{\rm{odd}}(t))=-\frac{dS_{\rm{odd}}(t)}{dt}.
\end{eqnarray}

Probability $S_{\rm{odd}}(t)$ is simply related to the conditional
probability $S(E_k,t)$ for a quasiparticle to occupy level $E_k$
at the moment $t$ in the box assuming that a quasiparticle entered
CPB at $t=0$ and resided continuously in the box during the time
interval $(0,t)$,
\begin{equation}\label{so}
S_{\rm{odd}}(t)=\sum_{k}S_{\rm{odd}}(E_k,t).
\end{equation}

We assume that in the initial moment of time the quasiparticle has
just entered the state $E_k$ in the box. Therefore, the initial
probability $S_{\textrm{o}}(E_k,0)$ of the occupation of the level
$E_k$ in the box is determined by the tunneling rate into the
state $E_k$,
\begin{eqnarray}\label{IC}
S_{\textrm{o}}(E_k,0)=\frac{1}{\Gamma_{\rm{in}}}\sum_{p}W(p,k)
f(E_p).
\end{eqnarray}
The normalization of $S_{\textrm{o}}(E_k,0)$ is chosen to satisfy
$S_{\rm{odd}}(0)=\sum_k S_{\textrm{o}}(E_k,0)=1$. According to
Eq.~(\ref{IC}) the initial conditional probability
$S_{\textrm{o}}(E_k,0)$ is zero below the threshold energy $E_k <
E_{\rm{thd}}$, and is proportional to Gibbs factor above the
threshold $E_k>E_{\rm{thd}}$. This reflects out-of-equilibrium
quasiparticle distribution at $t=0$.

The conditional probability $S_{\rm{odd}}(E_k,t)$ consistent with
initial conditions~(\ref{IC}) satisfies the following master
equation:
\begin{eqnarray}\label{master}
\!\dot{S}_{\textrm{o}}(E_k,t)\!+\!\Gamma_{\rm{out}}(E_k)S_{\textrm{o}}(E_k,t)=-\frac{S_{\rm{odd}}(E_k,t)\!-\!S^{\rm{eq}}_{\rm{odd}}(E_k,t)}{\tau}.\nonumber\\
\end{eqnarray}
The second term on the left-hand side corresponds to the loss from
the state $E_k$ due to the tunneling through the junction to the
lead with the rate $\Gamma_{\rm{out}}(E_k)$ of
Eq.~(\ref{gammaout}). Note that unlike in the theory of the rate
equations there is no ``gain" term in Eq.~(\ref{master}). This is
due to the condition that the box is occupied at $t=0$ and remains
occupied continuously until some time $t$. The right-hand side of
Eq.~(\ref{master}) corresponds to the electron-phonon collision
integral in the relaxation time approximation with $\tau$ of
Eq.~(\ref{tau}) and
$$S^{\rm{eq}}_{\rm{odd}}(E_k,t)=S_{\rm{odd}}(t)\cdot\rho^b_{\rm{odd}}(E_k).$$
Note that Eq.~(\ref{master}) is nonlocal in $E_k$ due to the
dependence of the collision integral on $S_{\rm{odd}}(t)$ (see
Eq.~(\ref{so})). The collision integral in Eq.~(\ref{master})
describes the phonon-induced relaxation of the trapped
quasiparticle to an equilibrium,
\begin{eqnarray}
\rho^b_{\rm{odd}}(E_k)=\frac{\exp\left(-
E_k/T\right)}{Z_{\rm{odd}}}.
\end{eqnarray}
Here $T$ is the quasiparticle temperature in the box. (For
simplicity, we assume that the effective quasiparticle temperature
in the lead is the same as in the box, $T_l=T_b=T$.) The
normalization factor $Z_{\rm{odd}}$  at $T\ll T^{*}$ is given by
\begin{eqnarray}\label{norm}
Z_{\rm{odd}}=\sqrt{2\pi}\frac{\Delta}{\delta_b}\sqrt{\frac{T}{\Delta}}\exp\left(-\frac{\Delta}{T}\right).
\end{eqnarray}

\subsection{General solution for $S_{\rm{odd}}(t)$.}

Using Laplace transform,
\begin{eqnarray}
S_{\textrm{o}}(E_k,s)=\int_0^{\infty}{dt}S_{\textrm{o}}(E_k,t)e^{-st},
\end{eqnarray}
we reduce the differential equation~(\ref{master}) supplied with
the initial conditions~(\ref{IC}) to an algebraic one
\begin{eqnarray}\label{inv_surv}
\left(\! s \!+\!
\Gamma_{\rm{out}}(E_k)\!+\!\frac{1}{\tau}\!\right)
\!S_{\textrm{o}}(E_k,s)
\!=\!\frac{S^{\rm{eq}}_{\rm{odd}}(E_k,s)}{\tau}\!+\!S_{\textrm{o}}(E_k,0).\nonumber\\
\end{eqnarray}
Equation~(\ref{inv_surv}) can be solved for
$S_{\textrm{o}}(E_k,s)$. Then, by summing that solution over
momenta $k$ and utilizing Eqs.~(\ref{so}) and~(\ref{IC}) we find
the survival probability $S_{\rm{odd}}(s)$
\begin{eqnarray}\label{sigma1}
S_{\rm{odd}}(s)=\frac{B(s)}{1-A(s)}.
\end{eqnarray}
Here functions $B(s)$ and $A(s)$ are defined as
\begin{eqnarray}\label{AandB}
B(s)&=&\frac{1}{\Gamma_{\rm{in}}}\sum_k\frac{f(E_k-\delta E)\Gamma_{\rm{out}}(E_k)}{s+1/\tau+\Gamma_{\rm{out}}(E_k)},\nonumber\\\\
A(s)&=&\frac{1}{\tau}\sum_k\frac{\rho^b_{\rm{odd}}(E_k)}{s+1/\tau+\Gamma_{\rm{out}}(E_k)}.\nonumber
\end{eqnarray}
At $T\gg\delta_b$ one can take the thermodynamic limit and replace
the sums with the integrals in Eq.~(\ref{AandB}). Further
simplification of the denominator in Eq.~(\ref{sigma1}) is
possible if one splits the integral in $A(s)$ into the intervals
$(\Delta,E_{\rm{thd}})$, where $\Gamma_{\rm{out}}(E_k)=0$, and
$(E_{\rm{thd}},\infty)$. Then, Eq.~(\ref{sigma1}) becomes
\begin{eqnarray}\label{sigma2}
S_{\rm{odd}}(s)=\left(s+\frac{1}{\tau}\right)\frac{B(s)}{s+X(s)}
\end{eqnarray}
with the functions $B(s)$ and $X(s)$ defined as
\begin{eqnarray}\label{BandX}
B(s)\!&\!=\!&\!\frac{2}{\Gamma_{\rm{in}}}\int_{E_{\rm{thd}}}^{\infty}\frac{d E_k}{\delta_b}\nu(E_k)\frac{f(E_k\!-\!\delta E)\Gamma_{\rm{odd}}(E_k)}{s\!+\!1/\tau\!+\!\Gamma_{\rm{odd}}(E_k)},\nonumber\\\nonumber\\
X(s)\!&\!=\!&\!\frac{2}{\tau}\int_{E_{\rm{thd}}}^{\infty}\frac{dE_k}{\delta_b}\nu(E_k)\frac{\rho^b_{\rm{odd}}(E_k)\Gamma_{\rm{odd}}(E_k)}{s\!+\!1/\tau\!+\!\Gamma_{\rm{odd}}(E_k)}.
\end{eqnarray}

\begin{figure}
\centering
\includegraphics[width=2.4in, scale=1]{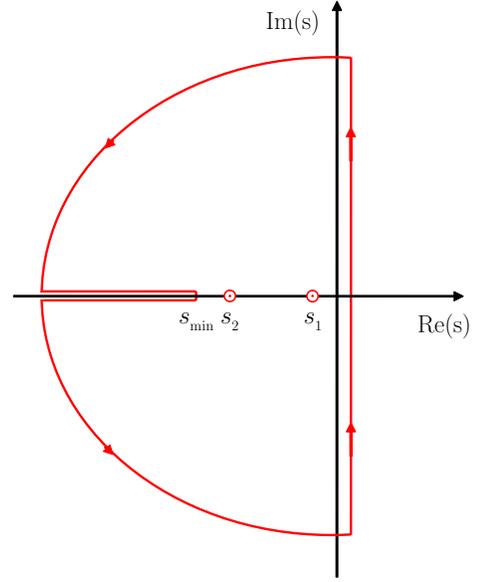}
\caption{(color online). Contour of integration (red line) chosen
to calculate inverse Laplace transform Eq.~(\ref{Brom}). Points of
non-analytic behavior of $\sigma_{_\textrm{++}}(s)$ are shown.
Poles at $s_1$, $s_2$, and a cut $s \in (-\infty,
-s_{\rm{min}}$).} \label{contour}
 \end{figure}

The inverse Laplace transform is given by
\begin{eqnarray}\label{Brom}
\!S_{\textrm{o}}(t)\!=\!\frac{1}{2\pi
i}\!\int_{\epsilon-i\infty}^{\epsilon+i\infty}{ds}\,S_{\rm{odd}}(s)e^{st},
\end{eqnarray}
where $\epsilon$ is chosen in such way that $S_{\rm{odd}}(s)$ is
analytic at $\mbox{\rm{Re}} [s]>\epsilon$. The
integral~(\ref{Brom}) can be calculated using complex variable
calculus by closing the contour of integration as shown in
Fig.~\ref{contour} and analyzing the enclosed points of
non-analytic behavior of $S_{\rm{odd}}(s)$. In general, the
singularities of $S_{\rm{odd}}(s)$ consist of two poles and a cut.
The latter is due to the singularities of the function $B(s)$
causing $S_{\rm{odd}}(s)$ to be non-analytic along the cut $s \in
\left(-\infty, -s_{\rm min} \right)$, where
\begin{eqnarray}
s_{\rm min}\!&\!=\!&\!\frac{1}{\tau}\!+\!\mbox{\rm min}
\left[\,\Gamma_{\rm{odd}}(E_k)\,\right].
\end{eqnarray}
The plot of $\Gamma_{\rm{odd}}(E_k)$ is shown in
Fig.~(\ref{fig_gammaout}). The function $\Gamma_{\rm{odd}}(E_k)$
has a minimum at $E_k^{\rm{min}}=E_{\rm{thd}}+\delta E/2$. (For
the estimate of the minimum we assumed $\delta E\ll \Delta$.) In
addition to the cut, $S_{\rm{odd}}(s)$ has 2 poles. The poles
$s_1$ and $s_2$ are the solutions of the following equation in the
region of analyticity of $B(s)$:
\begin{eqnarray}\label{poles}
s+X(s)=0.
\end{eqnarray}
We now analyze the singularities $S_{\rm{odd}}(s)$ and find their
contribution to the integral (\ref{Brom}).

The contribution from the cut to Eq.~(\ref{Brom}) corresponds to
the kinetics of unequilibrated quasiparticles. Formally it comes
from the non-analyticity of $S_{\rm{odd}}(s)$ due to the
singularities of the function $B(s)$ itself. The proper
contribution to Eq.~(\ref{Brom}) can be calculated by integrating
along the contour enclosing the cut,
\begin{eqnarray}\label{cut0}
\!I_{\rm cut}\!&\!=\!&\!\frac{\!-\!1}{2\pi i}\int_{s_{\rm
min}}^{\infty}{ds}e^{st}\left(S_{\rm{odd}}(s\!+\!i\epsilon)\!-\!S_{\rm{odd}}(s\!-\!i\epsilon)\right).\nonumber\\
\end{eqnarray}
At low temperature $T\ll \delta E$, the discontinuity of the
imaginary part of $S_{\rm{odd}}(s)$ at the cut is
\begin{eqnarray}\label{Impart3}
S_{\rm{odd}}(s\!+\!i\epsilon)\!-\!S_{\rm{odd}}(s\!-\!i\epsilon)=2i\left(s\!+\!\frac{1}{\tau}\right)\frac{
\mbox{\rm Im}{B(s\!+\!i\epsilon)}}{s}.
\end{eqnarray}
Substitution of this expression into Eq.~(\ref{cut0}) yields
\begin{eqnarray}
\!I_{\rm cut}
\!&\!=\!&\!\frac{2}{\Gamma_{\rm{in}}}\int_{E_{\rm{thd}}}^{\infty}\frac{dE_k}{\delta_b}\,\nu(E_k)f(E_k\!-\!\delta
E)\Gamma_{\rm{odd}}(E_k)\nonumber\\\nonumber\\
\!&\!\times\!&\!\frac{\tau\Gamma_{\rm{odd}}(E_k)}{1\!+\!\tau\Gamma_{\rm{odd}}(E_k)}\exp\!\left(-\frac{t}{\tau}-\Gamma_{\rm{odd}}(E_k)t\!\right)\!.
\end{eqnarray}
To simplify the above expression we introduce the dimensionless
variable $z$,
\begin{eqnarray}\label{z}
z=\frac{E_k-E_{\rm{thd}}}{T},
\end{eqnarray}
and write the integral in $I_{\rm cut}$ in terms of $z$,
\begin{eqnarray}\label{cut2}
\!I_{\rm
cut}\!&\!=\!&\!\frac{1}{\sqrt{\pi}\Gamma_{\rm{out}}\nu(\delta E)}\int_{0}^{\infty}dz\nu(z)\Gamma_{\rm{odd}}(z)\frac{\tau\Gamma_{\rm{odd}}(z)}{1+\tau\Gamma_{\rm{odd}}(z)}\nonumber\\\nonumber\\
&\!\times\!&\exp\left(-z-\Gamma_{\rm{odd}}(z)t-t/\tau\right).
\end{eqnarray}
Here and thereafter $\Gamma_{\rm{odd}}(z)$ and $\nu(z)$ are given
by Eqs.~(\ref{gammaout}) and~(\ref{DOS}), respectively, with
$E_k=E_{\rm{thd}}+T z$.

We now analyze the contribution to Eq.~(\ref{Brom}) from the
poles. The pole at $s_1$ may be found from the iterative solution
of Eq.~(\ref{poles}) at small $s$ ($s\ll s_{\rm min}$),
\begin{eqnarray}\label{s1}
s_1\approx -X(s=0),
\end{eqnarray}
with $X(s)$ given by Eq.~(\ref{BandX}). The contribution from the
pole at $s_1$ can be easily calculated using residue calculus
yielding
\begin{eqnarray}\label{I1}
I_{1}\!&\!=\!&\!Y(0)\exp\left[-X(0)t\right].
\end{eqnarray}
Equation~(\ref{I1}) describes the kinetics of thermalized
quasiparticles. At low temperature $X(0)\propto \exp(-\delta
E/T)$, which justifies the approximation used in Eq.~(\ref{s1}),
see also next section. The function $Y(0)$ depends on
$\tau\Gamma_{\rm{out}}$, and is approximately given by
\begin{eqnarray}\label{Y0}
Y(0)\approx\frac{1}{\sqrt{\pi}}\int_{0}^{\infty} d
z\frac{\exp\left(-z\right)}{\sqrt{z}+\tau \Gamma_{\rm{out}}}.
\end{eqnarray}
Here we used small-$z$ asymptote ($z\ll\frac{\delta E}{2T}$) for
the escape rate,
\begin{eqnarray}\label{smallz}
\Gamma_{\rm{out}}(z)\approx\frac{\Gamma_{\rm{out}}}{\sqrt{z}}.
\end{eqnarray}
The second pole $s_2$ is given by the solution of
Eq.~(\ref{poles}) at large $s$. At small temperature $T\ll\delta
E$ one can show that the contribution of the second pole $s_2$ to
Eq.~(\ref{Brom}) is small, and thus can be neglected. (For
details, see the Appendix in Ref.~[\onlinecite{Lutchyn2}] )

\subsection{Results and Discussions.}

Combining all relevant contributions to the inverse Laplace
transform, Eqs.~(\ref{cut2}) and~(\ref{I1}), we obtain the
solution for the survival probability
\begin{eqnarray}\label{surv}
S_{\rm{odd}}(t)&=&Y(0)\exp\left(-\gamma t \right)+F(t).
\end{eqnarray}
The first term here corresponds to the kinetics of the
quasiparticle that relaxed to the bottom of the trap. The
thermally activated decay rate $\gamma$, found with the help of
Eqs.~(\ref{s1}) and~(\ref{z}), is
\begin{eqnarray}\label{gamma}
\!\gamma\!=\!\frac{1}{\tau}\frac{\nu(\delta
E)}{\nu(T)}\exp\!\left(-\frac{\delta
E}{T}\right)\!\left(1\!-\!\int_0^{\infty}\!dz\frac{e^{-z}/\tau }{1/\tau+\Gamma_{\rm{out}}/\sqrt{z}}\!\right)\!.\nonumber\\
\end{eqnarray}
The integral in Eq.~(\ref{gamma}) reflects the probability for a
quasiparticle to relax to the bottom of the trap [cf.
Eq.~(\ref{Ptr})]. The second term in Eq.~(\ref{surv}) describes
the kinetics of unequilibrated quasiparticles with $F(t)$ given by
\begin{eqnarray}\label{Function_t}
F(t)\!&\!=\!&\!\frac{1}{\sqrt{\pi}\Gamma_{\rm{out}}\nu(\delta E)}\int_{0}^{\infty}dz\nu(z)\Gamma_{\rm{odd}}(z)\frac{\tau\Gamma_{\rm{odd}}(z)}{1+\tau\Gamma_{\rm{odd}}(z)}\nonumber\\\nonumber\\
\!&\!\times\!&\!\exp\left(-z-t\Gamma_{\rm{odd}}(z)-\frac{t}{\tau}\right).
\end{eqnarray}
In the next paragraphs we will analyze $S_{\rm{odd}}(t)$ for fast
and slow relaxation limits.

In the fast relaxation limit $\tau \Gamma_{\rm{out}} \ll 1$
(``deep" trap), the leading contribution to the survival
probability $S_{\rm{odd}}(t)$ comes from the first term in
Eq.~(\ref{surv}), the second term in Eq.~(\ref{surv}) is
proportional to $~\tau \Gamma_{\rm{out}}$, and can be neglected.
Consequently, the survival probability is given by
\begin{eqnarray}\label{fast_asy}
S_{\rm{odd}}(t)&\approx&\exp\left(-\gamma_{{\rm{f}}} t\right).
\end{eqnarray}
Using Eq.~(\ref{dwell_distribution}) we find the lifetime
distribution function
\begin{eqnarray}
N_{\rm{odd}}(t)=\gamma_{{\rm{f}}}\exp\left(-\gamma_{{\rm{f}}}
t\right),
\end{eqnarray}
cf. Eqs.~(\ref{tail}) and~(\ref{qual_rate_fast}). As discussed
qualitatively in Sec.~\ref{theomodel} in the fast relaxation limit
the majority of quasiparticles entering the CPB into the excited
state $E_k\sim E_{\rm{thd}}$ relax to the bottom of the trap and
stay in the box until they are thermally excited out of the trap
by phonons  with the rate $\gamma_{\rm{f}}$ of
Eq.~(\ref{qual_rate_fast}). Finally, using Eq.~(\ref{fast_asy}) we
find the average lifetime of the odd-charge state
\begin{eqnarray}\label{to}
\langle T_{\rm{odd}} \rangle &=&\int_0^{\infty} S_{\rm{odd}}(t)
dt=1/\gamma_{\rm{f}}.
\end{eqnarray}

In the opposite limit of the ``shallow" trap, $\tau
\Gamma_{\rm{out}}\gg 1$, the majority of quasiparticles tunnel out
unequilibrated to the lead ($P_{\rm{tr}}\approx
1/\tau\Gamma_{\rm{out}}$). The expression for the survival
probability~(\ref{surv}) in this limit becomes
\begin{eqnarray}\label{sigma_slow}
S_{\rm{odd}}(t)=F(t)+\frac{1}{\sqrt{\pi}\tau\Gamma_{\rm{out}}}\exp(-\gamma_{\rm{s}}
t).
\end{eqnarray}
Note that in addition to the first term describing the kinetics of
unequilibrated quasiparticles the survival probability has a tail
corresponding to the small fraction of quasiparticles that do
relax to the bottom of the trap.  These quasiparticles reside in
the box until they are thermally excited by phonons. In the slow
relaxation limit the activation exponent~(\ref{gamma}) can be
reduced to
\begin{eqnarray}\label{gamma_s}
\gamma_{\rm{s}}\approx\frac{1}{\sqrt{\pi}\tau}\frac{\nu(\delta
E)}{\nu(T)}\exp\left(-\frac{\delta E}{T}\right).
\end{eqnarray}
[Rigorous evaluation produces a difference in the numerical factor
here compared to Eq.~(\ref{qual_rate_slow}).] The tail of the
distribution function~(\ref{sigma_slow}) describes the processes
that are much slower than $1/\Gamma_{\rm{out}}$, thus it must be
retained despite its small amplitude, see also
Eq.~(\ref{todd_slow}).

\begin{figure}
\includegraphics[width=3.2in]{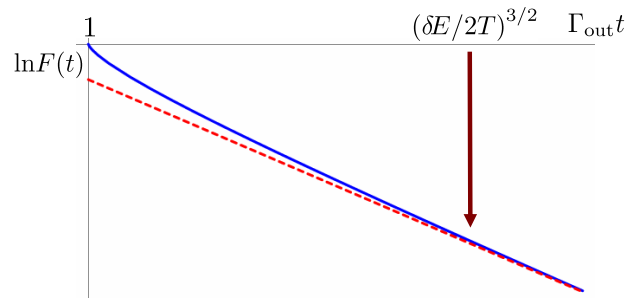}
\caption{(color online). Deviation of $F(t)$ (solid blue line)
defined in Eq.~(\ref{Function_t}) from the exponentially decaying
function at $\Gamma_{\rm{out}}t \gtrsim 1$. (We assumed
$\tau=\infty$ here.)
  } \label{Ft}
 \end{figure}

The function $F(t)$ defined in Eq.~(\ref{Function_t}) can be
evaluated using the small-$z$ asymptote of $\Gamma_{\rm{odd}}(z)$,
see Eq.~(\ref{smallz}). This approximation substantially
simplifies $F(t)$,
\begin{eqnarray}\label{Function_t_appr}
F(t)\approx\frac{1}{\sqrt{\pi}}\int_{0}^{\infty}\frac{dz}{\sqrt{z}}\frac{\tau
\Gamma_{\rm{out}}}{\sqrt{z}+\tau\Gamma_{\rm{out}}}\exp\!\left(\!-z\!-\!\frac{t\Gamma_{\rm{out}}}{\sqrt{z}}\!-\frac{t}{\tau}\!\right)\!.\nonumber\\
\end{eqnarray}
Here we assumed that the main contribution to the $F(t)$ comes
from small-$z$ region, $z\ll \delta E/2T$, which limits the
applicability of Eq.~(\ref{Function_t_appr}) to
$t\ll\Gamma_{\rm{out}}^{-1}\left(\frac{\delta
E}{2T}\right)^{3/2}$. The asymptotic expression for $F(t)$ in
Eq.~(\ref{sigma_slow}) can be obtained using the saddle-point
approximation
\begin{eqnarray}\label{Function_t_2}
\!F(t)\!\approx\!\frac{2}{\sqrt{3}}\frac{\tau
\Gamma_{\rm{out}}}{\tau
\Gamma_{\rm{out}}\!+\!\left(\frac{1}{2}\Gamma_{\rm{out}}t\right)^{1/3}}\exp\!\left(\!-3\!\left(\!\frac{\Gamma_{\rm{out}}t}{2}\!\right)^{2/3}\!-\frac{t}{\tau}\!\right)\!.\nonumber\\
\end{eqnarray}
The integral~(\ref{Function_t_appr}) can be also expressed in the
analytic form in terms of the Meijer's G-function~\cite{Stegun}.
As one can see from Fig.~\ref{Ft} at low temperature $T\ll\delta
E$ there is a time window
\begin{eqnarray}\label{timewindow}
\frac{1}{\Gamma_{\rm{out}}}\lesssim t \ll
\frac{1}{\Gamma_{\rm{out}}}\left(\frac{\delta E}{2T}\right)^{3/2},
\end{eqnarray}
in which the survival probability deviates from the exponentially
decaying function. We assumed in Eq.~(\ref{timewindow}) that the
upper limit is more restrictive than
$t\ll\frac{1}{\Gamma_{\rm{out}}}\left(\Gamma_{\rm{out}}\tau\right)^{3}$
so that $\tau$-dependent term in the exponent of
Eq.~(\ref{Function_t_2}) can be neglected.

The fractional power $2/3$ in Eq.~(\ref{Function_t_2}) stems from
the peculiarity of superconducting density of states at low
energies. Assuming the quasiparticle distribution in the lead is
given by Eq.~(\ref{distr_func}), every time a quasiparticle
tunnels into the box it may occupy a different energy level, which
is reflected in the initial conditions, Eq.~(\ref{IC}). However,
due to the singularity of the escape rate $\Gamma_{\rm{out}}(E_k)$
at $E_k\sim E_{\rm{thd}}$, this results in a strong energy
dependence of the dwell time of a quasiparticle. Therefore,
averaging over many such events leads to the deviation of $F(t)$
from the simple exponential function, as shown in Fig.~\ref{Ft}.

At $t\gtrsim\frac{1}{\Gamma_{\rm{out}}}\left(\frac{\delta
E}{2T}\right)^{3/2}$ the minimum of the exponent
in~(\ref{Function_t_appr}) is beyond the limit of applicability of
the small-$z$ approximation for the rate $\Gamma_{\rm{out}}(z)$
given by Eq.~(\ref{smallz}), and instead of
Eq.~(\ref{Function_t_appr}) one should use Eq.~(\ref{Function_t}).
Since at $z\sim \delta E/2T$ the escape rate
$\Gamma_{\rm{out}}(z)$ is a smooth function, $F(t)$ decays
exponentially,
\begin{eqnarray}\label{Function_t_3}
F(t)\!\propto\exp\!\left(-\frac{\delta
E}{2T}-\Gamma_{\rm{out}}(z_{\rm{min}})t-\frac{t}{\tau}\right)\!.
\end{eqnarray}
Here $\Gamma_{\rm{out}}(z_{\rm{min}})\approx
\frac{g_{_{\rm{T}}}\delta_b}{2\pi}\sqrt{\frac{\delta E}{\Delta}}$.

The lifetime distribution function $N_{\rm{odd}}(t)$ for the
odd-charge state can be obtained from $S_{\rm{odd}}(t)$ by
substituting Eq.~(\ref{sigma_slow}) into
Eq.~(\ref{dwell_distribution}). Under conditions of
Eq.~(\ref{timewindow}) the lifetime distribution function
$N_{\rm{odd}}(t)$ will deviate from the exponential distribution
\begin{eqnarray}
N_{\rm{odd}}(t)\approx\frac{2^{4/3}}{\sqrt{3}}\frac{\Gamma_{\rm{out}}}{\left(\Gamma_{\rm{out}}t\right)^{1/3}}\exp\!\left(-3\!\left(\!\frac{\Gamma_{\rm{out}}t}{2}\!\right)^{2/3}\right).
\end{eqnarray}

The average lifetime of the odd-charge state $\langle T_{\rm{odd}}
\rangle$ in the slow relaxation case is
\begin{eqnarray}\label{todd_slow}
\!\langle T_{\rm{odd}} \rangle\!=\!\int_0^{\infty}
\!S_{\rm{odd}}(t)
dt\!&\!\approx\!&\!\frac{1}{\sqrt{\pi}\tau\Gamma_{\rm{out}}\gamma_s}=\frac{1}{\gamma_f}.
\end{eqnarray}
Despite the quasiparticle having small probability of relaxing to
the bottom of the trap, the main contribution to the average dwell
time $\langle T_{\rm{odd}} \rangle$ is given by the tail of
$S_{\rm{odd}}(t)$. This is because once the quasiparticle is
trapped in the CPB it spends there exponentially long time, see
Eq.~(\ref{gamma_s}). As expected $\langle T_{\rm{odd}} \rangle$ is
the same for fast and slow relaxation cases since the average
lifetime determines the thermodynamic probability to occupy a
given charge state.

\section{charge noise.}\label{secnoise}

The complex statistics of capture and emission processes discussed
in the preceding section also manifest itself in the spectral
density of charge fluctuations of the Cooper-pair box. In this
section we study the charge noise power spectrum for ``deep"
($\tau\Gamma_{\rm{out}}\ll 1$) and ``shallow"
($\tau\Gamma_{\rm{out}}\gg~1$) traps.

The kinetic equations for occupational probabilities of odd- and
even-charge state have the form~\cite{Lutchyn2}
\begin{eqnarray}\label{KE}
\dot{P}_{\rm{even}}(E_p,t)\!&\!+\!&\!\sum_kW(p,k)\left(P_{\rm{even}}(E_p,t)\!-\!P_{\rm{odd}}(E_k,t)\right)\!=\!0,\nonumber\\
\dot{P}_{\rm{odd}}(E_k,t)\!&\!+\!&\!\sum_pW(p,k)\left(P_{\rm{odd}}(E_k,t)\!-\!P_{\rm{even}}(E_p,t)\right)\!=\!\nonumber\\
\!&\!-\!&\!\frac{1}{\tau}\left(P_{\rm{odd}}(E_k,t)-P^{\rm{eq}}_{\rm{odd}}(E_k,t)\right).
\end{eqnarray}
Here
$P^{\rm{eq}}_{\rm{odd}}(E_k,t)=\rho^b_{\rm{odd}}(E_k)\sigma_{\rm{odd}}(t)$
with $\sigma_{\rm{odd}}(t)=\sum_k P_{\rm{odd}}(E_k,t)$ , and the
quasiparticle transition rate $W(p,k)$ is defined in
Eq.~(\ref{rates2}). Assuming that the lead is a heat bath of
quasiparticles we can write even-charge occupational probability
as $P_{\rm{even}}(E_p,t)=f(E_p)\sigma_{\rm{even}}(t)$ with
$f(E_p)$ being the distribution function of the quasiparticles in
the lead, and $\sigma_{\rm{even}}(t)=\sum_p P_{\rm{even}}(E_p,t)$
being the occupational probability of the even state. This allows
us to reduce Eqs.~(\ref{KE}) to
\begin{eqnarray}\label{KE1}
\dot{\sigma}_{\rm{even}}(t)\!&\!+\!&\!\sum_{k,p}W(p,k)\left(f(E_p)\sigma_{\rm{even}}(t)\!-\!P_{\rm{odd}}(E_k,t)\right)\!=\!0\nonumber\\
\dot{P}_{\rm{odd}}(E_k,t)\!&\!+\!&\!\sum_pW(p,k)\left(P_{\rm{odd}}(E_k,t)\!-\!f(E_p)\sigma_{\rm{even}}(t)\right)\!=\!\nonumber\\
\!&\!-\!&\!\frac{1}{\tau}\left(P_{\rm{odd}}(E_k,t)\!-\!P^{\rm{eq}}_{\rm{odd}}(E_k,t)\right).
\end{eqnarray}
One can see that Eqs.~(\ref{KE1}) satisfy the normalization
condition:
\begin{eqnarray}\sigma_{\textrm{e}}(t)+\sigma_{\textrm{o}}(t)=1.
\end{eqnarray}
The stationary occupational probabilities
$\bar{\sigma}_{\rm{even}}$ and $\bar{\sigma}_{\rm{odd}}$ are given
by the Gibbs equilibrium state. Assuming that $f(E_p)$ is given by
Eq.~(\ref{distr_func}), we obtain
\begin{eqnarray}\label{stationary1}
\bar{\sigma}_{\rm{even}}=\frac{1}{1+n_{\rm{qp}}^lV_b\exp\left(
\frac{\delta E}{T}\right)}\mbox{, }
\bar{\sigma}_{\rm{odd}}=1-\bar{\sigma}_{\rm{even}}.
\end{eqnarray}
Here  $n_{\rm{qp}}^l$ is the quasiparticle density in the lead,
see Eq.~(\ref{density}), and $V_b$ is the volume of the CPB.

The fluctuations around this equilibrium state can be taken into
account within the Boltzmann-Langevin approach, which assumes that
the occupational probabilities fluctuate around the stationary
solution~(\ref{stationary1}) due to the randomness of the
tunneling and scattering events as well as partial occupations of
the quasiparticle states~\cite{footnote2}. The kinetic equations
for the charge fluctuations can be derived by properly varying
Eqs.~(\ref{KE1}) and adding Langevin sources corresponding to the
relevant random events~\cite{Kogan},
\begin{eqnarray}\label{BLE}
\!&\!\,\!&\!\left(\frac{d}{dt}\!+\!\Gamma_{\rm{in}}\right)\delta\sigma_{\rm{even}}(t)\!=\!\sum_{k,p}W(p,k)\delta P_{\rm{odd}}(E_k,t)\!+\!\sum_p\xi_p^{_{\rm{T}}}(t),\nonumber\\
\!&\!\,\!&\!\left(\partial_t\!+\!\sum_pW(p,k)\!+\!\frac{1}{\tau}\right)\delta
P_{\rm{odd}}(E_k,t)\!=\!-\frac{\delta
\sigma_{\rm{even}}(t)}{\tau}\rho^b_{\rm{odd}}(E_k)\nonumber\\
\!&\!\,\!&\!+\!\sum_pW(p,k)f(E_p)\delta\sigma_{\rm{even}}(t)\!+\!\xi_k^{_{\rm{T}}}(t)\!+\!\xi_k^{_{ph}}(t).
\end{eqnarray}
Here the relation $\delta \sigma_{\rm{even}}(t)=-\delta
\sigma_{\rm{odd}}(t)$ was taken into account. The Langevin sources
$\xi_{p(k)}^{_{\rm{T}}}(t)$ and $\xi_k^{_{ph}}(t)$ correspond to
quasiparticle tunneling from/to the state $\ket{p}/\ket{k}$
through the junction, and inelastic electron-phonon scattering,
respectively. [Note that $\sum_p\xi_{p}^{_{\rm{T}}}(t)=-\sum_k
\xi_{k}^{_{\rm{T}}}(t)$ and $\sum_k \xi_k^{_{ph}}(t)=0$.] These
random processes are considered to be Poissonian with the
following correlation functions
\begin{eqnarray}\label{correlations}
\!\langle\xi_k^{_{\rm{T}}}(t)\xi_{k'}^{_{\rm{T}}}(t')\rangle
\!&\!=\!&\! 2\delta(t-t')\delta_{k,k'}\sum_p
W(p,k)f(E_p)\bar{\sigma}_{\rm{even}}\nonumber\\
\!&\!=\!&\!2\delta(t-t')\delta_{k,k'}\Gamma_{\rm{out}}(E_k)f(E_k\!-\!\delta E)\bar{\sigma}_{\rm{even}},\nonumber\\\nonumber\\
\!\langle\xi_k^{_{ph}}(t)\xi_{k'}^{_{ph}}(t')\rangle \!&\!=\!&\!
\delta(t\!-\!t')\frac{2P^{\rm{eq}}_{\rm{odd}}(E_k)}{\tau}\left(\delta_{k,k'}\!-\!\frac{P^{\rm{eq}}_{\rm{odd}}(E_{k'})}{\sigma_{\rm{odd}}}\right)\nonumber\\
\!&\!=\!&\!\delta(t\!-\!t')\frac{2\bar{\sigma}_{\rm{odd}}\rho^b_{\rm{odd}}(E_k)}{\tau}\left(\delta_{k,k'}\!-\!\rho^b_{\rm{odd}}(E_{k'})\right).\nonumber\\
\end{eqnarray}
The latter expression is consistent with the collision integral in
the relaxation time approximation and conservation of the
probability $\sigma_{\rm{odd}}(t)$ by the electron-phonon
scattering~\cite{Lax, Zwanzig}.

The spectral density of charge fluctuations in the CPB is defined
as
\begin{eqnarray}\label{noise_def}
S_Q(\omega)=2e^2\langle\delta\sigma_{\rm{even}}(\omega)\delta\sigma_{\rm{even}}(-\omega)\rangle,
\end{eqnarray}
and can be obtained from Eqs.~(\ref{BLE})
and~(\ref{correlations}). The solution of the second equation of
the system~(\ref{BLE}) in frequency domain is
\begin{eqnarray}\label{sol1}
\delta
P_{\rm{odd}}(E_k,\omega)&=&\frac{\Gamma_{\rm{out}}(E_k)f(E_k\!-\!\delta
E)-\frac{1}{\tau}\rho^b_{\rm{odd}}(E_k)}{-i\omega+
\Gamma_{\rm{out}}(E_k)+\frac{1}{\tau}}\delta
\sigma_{\rm{even}}(\omega)\nonumber\\
&+&\frac{\xi_k^{_{\rm{T}}}(\omega)+\xi_k^{_{ph}}(\omega)}{-i\omega+\Gamma_{\rm{out}}(E_k)+\frac{1}{\tau}}.
\end{eqnarray}
Substituting this expression into the equation for $\delta
\sigma_{\rm{even}}(\omega)$ we find

\begin{eqnarray}\label{sol2}
\mathcal{L}(\omega)\delta\sigma_{\rm{even}}(\omega)\!=\!\sum_{k}\frac{(i\omega\!-\!\frac{1}{\tau})\xi_k^{_{\rm{T}}}(\omega)\!+\!\Gamma_{\rm{out}}(E_k)\xi_k^{_{ph}}(\omega)}{-i\omega+\Gamma_{\rm{out}}(E_k)+\frac{1}{\tau}},
\end{eqnarray}
where the function $\mathcal{L}(\omega)$ is given by
\begin{eqnarray}\label{L}
\mathcal{L}(\omega)&=&-i\omega+\frac{1}{\tau}\sum_k\frac{\rho^b_{\rm{odd}}(E_k)\Gamma_{\rm{out}}(E_k)}{-i\omega+\Gamma_{\rm{out}}(E_k)+\frac{1}{\tau}}\nonumber\\
&+&\sum_{k}f(E_{k}\!-\!\delta
E)\frac{\Gamma_{\rm{out}}(E_k)(-i\omega+\frac{1}{\tau})}{-i\omega+\Gamma_{\rm{out}}(E_k)+\frac{1}{\tau}}.
\end{eqnarray}

Finally, using Eqs.~(\ref{noise_def}) and~(\ref{sol2}) we can find
the correlation function
$\langle\delta\sigma_{\rm{even}}(\omega)\delta\sigma_{\rm{even}}(-\omega)\rangle$,
and obtain charge noise power spectrum $S_Q(\omega)$,

\begin{widetext}
\begin{eqnarray}\label{main_noise}
S_Q(\omega)=\frac{2e^2}{\mathcal{L}(\omega)\mathcal{L}(-\omega)}\sum_{k,k'}\frac{(\omega^2\!+\!\frac{1}{\tau^2})\langle\xi_k^{_{\rm{T}}}(\omega)\xi_{k'}^{_{\rm{T}}}(-\omega)\rangle\!+\!\Gamma_{\rm{out}}(E_k)\Gamma_{\rm{out}}(E_{k'})\langle\xi_k^{_{ph}}(\omega)\xi_{k'}^{_{ph}}(-\omega)\rangle}{\left(-i\omega+\Gamma_{\rm{out}}(E_k)+\frac{1}{\tau}\right)\left(i\omega+\Gamma_{\rm{out}}(E_{k'})+\frac{1}{\tau}\right)}.
\end{eqnarray}
\end{widetext}
Upon substituting correlation functions~(\ref{correlations}) into
Eq.~(\ref{main_noise}) the general solution for $S_Q(\omega)$ can
be obtained (after cumbersome but straightforward calculations,
see the Appendix). Rather than going through the full derivation,
we study here $S_Q(\omega)$ in the limiting cases
$\tau\Gamma_{\rm{out}}\ll 1$ and $\tau\Gamma_{\rm{out}}\gg 1$,
which can be derived from Eqs.~(\ref{correlations}), (\ref{L}),
and (\ref{main_noise}).

We first consider fast relaxation limit $\tau\Gamma_{\rm{out}}\ll
1$. In this case one can neglect the second term in the numerator
of Eq.~(\ref{main_noise}). For $\omega\tau\ll 1$ one can simplify
Eqs.~(\ref{L}) and~(\ref{main_noise}) further. After
straightforward manipulations one finds that the leading
contribution to the noise is given by Eq.~(\ref{qual_noise_fast})
with the rate
\begin{eqnarray}\label{width_fast}
\frac{1}{\tau_{\rm{eff}}}=\gamma_{\rm{f}}+\Gamma_{\rm{in}},
\end{eqnarray}
which includes all processes changing the population
$\bar{\sigma}_{\rm{even}}$. The first term in
Eq.~(\ref{width_fast}) corresponds to the rate of thermal
activation of the quasiparticles by phonons from the bottom of the
trap to the lead, see Eq.~(\ref{qual_rate_fast}); the second term
is the rate of quasiparticle tunneling from the lead to the box
given by Eq.~(\ref{Gamma_in}), also cf.
Eq.~(\ref{qual_width_fast}).

\begin{figure}
\centering
\includegraphics[width=3.4in]{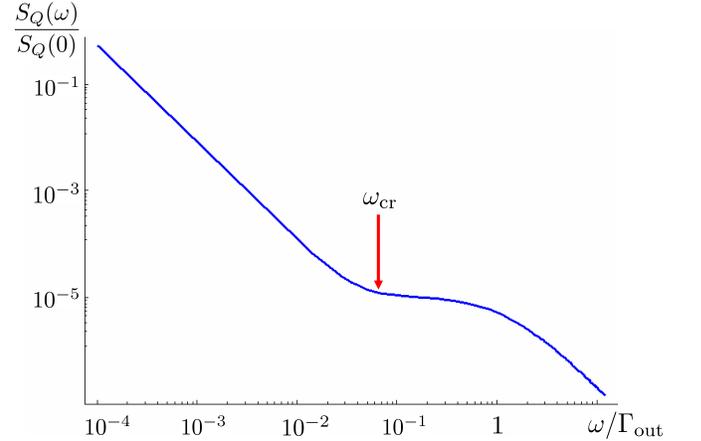}
\caption{(color online). Spectral density of charge fluctuations
generated by quasiparticle capture and emission processes in the
Cooper-pair box for the slow relaxation case ($\tau
\Gamma_{\rm{out}}=10^3$). Here $\omega_{\rm{cr}}\approx
\sqrt{\Gamma_{\rm{out}}/\tau}$ is a crossover frequency between
two different regimes governed by Eqs.~(\ref{noise_slow_1})
and~(\ref{noise_slow1}).  } \label{fig_noise_slow}
\end{figure}

In the opposite limit $\tau\Gamma_{\rm{out}}\gg 1$, the charge
noise power spectrum $S_Q(\omega)$ can be roughly viewed as the
superposition of two Lorentzians, see Fig.~\ref{fig_noise_slow}.
The first one corresponds to the processes involving quasiparticle
thermalization and activation by phonons, and is dominant at low
frequencies. The second (quasi) Lorentzian describes the fast
processes ($\omega\!\sim\! \Gamma_{\rm{out}}$) associated with the
escape of unequilibrated quasiparticles from the box.

At low frequencies $\omega\ll \omega_{\rm{cr}}$, see
Fig.~\ref{fig_noise_slow}, the noise power spectrum is well
approximated by the Lorentzian function. This can be obtained  by
neglecting the first term in the numerator of
Eq.~(\ref{main_noise}), and keeping the leading terms in
$1/\tau\Gamma_{\rm{out}}$ and $\omega/\Gamma_{\rm{out}}$ in
Eqs.~(\ref{L}) and~(\ref{main_noise}), see the Appendix. After
straightforward manipulations one finds
\begin{eqnarray}\label{noise_slow_1}
S_Q(\omega)\approx4e^2\bar{\sigma}_{\rm{odd}}\left(1-\bar{\sigma}_{\rm{odd}}\right)\frac{1-D}{1+C}\cdot\frac{\tau_{\rm{eff}}}{(\omega\tau_{\rm{eff}})^2+1}.
\end{eqnarray}
The constants $C$ and $D$ here are defined as
\begin{eqnarray}\label{CandD}
C=\frac{1}{\sqrt{\pi}}\frac{\Gamma_{\rm{in}}}{\Gamma_{\rm{out}}},\mbox
{ and } D=\frac{1}{\sqrt{\pi}}\frac{\nu(\delta
E)}{\nu(T)}\exp\left(-\frac{\delta E}{T}\right).
\end{eqnarray}
The width of the Lorentzian~(\ref{noise_slow_1})  is given by
\begin{eqnarray}\label{width_slow}
\frac{1}{\tau_{\rm{eff}}}\!=\!\frac{1}{\tau}\frac{\Gamma_{\rm{in}}}{\Gamma_{\rm{in}}\!+\!\sqrt{\pi}\Gamma_{\rm{out}}}\!+\!\gamma_{\rm{s}}\frac{\sqrt{\pi}\Gamma_{\rm{out}}}{\Gamma_{\rm{in}}\!+\!\sqrt{\pi}\Gamma_{\rm{out}}}.
\end{eqnarray}
The first term here corresponds to the transitions from even- to
odd-charge state involving the relaxation of a quasiparticle to
the bottom of the trap. [cf. Eq.~(\ref{qual_noise_slow1});
difference in the numerical coefficients comes from rigorous
solution of Eqs.~(\ref{correlations}), (\ref{L}) and
(\ref{main_noise}).] It is determined by the quasiparticle
relaxation rate $1/\tau$ times the portion of the time the
unequilibrated quasiparticle spends in the box. The second term in
Eq.~(\ref{width_slow}) describes the transitions odd to even state
involving the escapes of a thermalized quasiparticle from the CPB
by phonon activation. It is proportional to the phonon-assisted
quasiparticle escape rate from the box to the lead $\gamma_s$
times the probability to find an empty trap upon the escape of the
thermalized quasiparticle. This probability is determined by the
portion of the time the trap spends in the even state upon the
escape of the thermalized quasiparticle, and is determined by the
fast processes involving $\Gamma_{\rm{out}}$ and
$\Gamma_{\rm{in}}$.

At high frequencies, $\omega\gg \omega_{\rm{cr}}$, the dominant is
the first term in the numerator of Eq.~(\ref{main_noise}). Then,
in the leading order in $1/\tau\Gamma_{\rm{out}}$ the power
spectrum becomes
\begin{eqnarray}\label{noise_slow1}
\!S_Q(\omega)\!\approx\!\frac{4e^2}{\Gamma_{\rm{out}}}\!\cdot\!\!\frac{C
Z_1(\omega)\bar{\sigma}_{\rm{even}}}{\left(1\!+\!C
Z_2(\omega)\right)^2\!+\!\left(\frac{\omega}{\Gamma_{\rm{out}}}\right)^2\left(C
Z_1(\omega)\right)^2}.
\end{eqnarray}
Here the sums over momentum $k$ in Eq.~(\ref{main_noise}) are
replaced with the integrals ($T\gg \delta_b$). In terms of the
dimensionless variable $z$~(\ref{z}) these integrals are denoted
as $Z_1(\omega)$ and $Z_2(\omega)$ [see the Appendix],
\begin{eqnarray}\label{z1_2}
Z_1(\omega)\!\approx\int_0^{\infty}dz
\frac{e^{-z}\sqrt{z}}{\left(\omega/\Gamma_{\rm{out}}\right)^2z+1},\nonumber\\\nonumber\\
Z_2(\omega)\!\approx\int_0^{\infty}dz
\frac{e^{-z}}{\left(\omega/\Gamma_{\rm{out}}\right)^2z+1}.
\end{eqnarray}
As shown in Fig.~\ref{fig_noise_slow}, in the frequency window
$\omega_{\rm{cr}}\ll\omega\ll \Gamma_{\rm{out}}$ the noise power
$S_Q(\omega)$ becomes flat with the amplitude
\begin{eqnarray}
S_Q(\omega)\approx\frac{2\sqrt{\pi}e^2}{\Gamma_{\rm{out}}}\frac{C\bar{\sigma}_{\rm{even}}}{(1+C)^2}.
\end{eqnarray}
\begin{figure}
\centering
\includegraphics[width=3.2in]{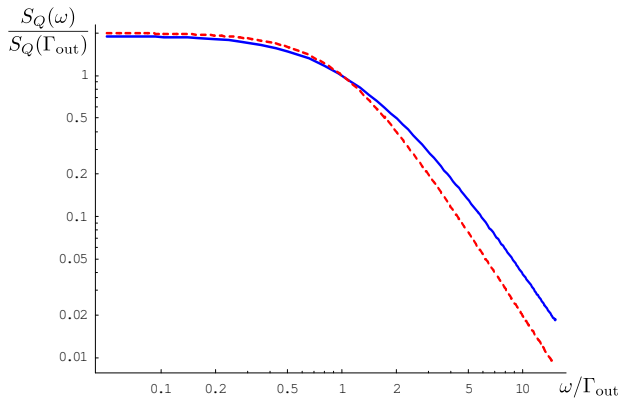}
\caption{(color online). The deviations of the charge noise power
spectrum $S_Q(\omega)$ from the Lorentzian function at high
frequencies $\omega\!\sim\!\Gamma_{\rm{out}}$. Blue solid line
corresponds to $S_Q(\omega)$ given by Eq.~(\ref{noise_slow_dev}),
red dashed line is the normalized Lorentzian function with the
width $\Gamma_{\rm{out}}$.} \label{deviations}
\end{figure}
At higher frequencies $\omega\gtrsim \Gamma_{\rm{out}}$ and $C\ll
1$ the noise power spectrum~(\ref{noise_slow1}) can be
approximated by
\begin{eqnarray}\label{noise_slow_dev}
\!S_Q(\omega)\!\approx\!\frac{4e^2}{\Gamma_{\rm{out}}} C
Z_1(\omega)\bar{\sigma}_{\rm{even}}
\end{eqnarray}
with $Z_1(\omega)$ given by Eq.~(\ref{z1_2}). At these frequencies
the charge noise power spectrum $S_Q(\omega)$ describes charge
fluctuations due to the tunneling of the unequilibrated
quasiparticles from the box to the lead. By taking a Fourier
transform of Eq.~(\ref{noise_slow_dev}), one can notice that the
noise power spectrum in time domain has the same functional form
as $F(t)$ of Eq.~(\ref{Function_t_2}). Therefore, charge noise
power spectrum also reveals the deviations from the conventional
Poisson statistics due to the singularity of the quasiparticle
density of states at low energies. The deviations of the charge
noise power spectrum~(\ref{noise_slow_dev}) from the Lorentzian
function at $\omega\!\sim\!\Gamma_{\rm{out}}$ are illustrated in
Fig.~(\ref{deviations}). At higher frequencies $\omega\gg
\Gamma_{\rm{out}}$ charge noise power spectrum $S_Q(\omega)$
decays as $1/\omega^2$, see Eq.~(\ref{qual_noise_slow2}).

\section{Conclusions.}\label{conclusion}

In this work we studied the kinetics of the quasiparticle trapping
and releasing in the mesoscopic superconducting island
(Cooper-pair box). We found the lifetime distribution of even- and
odd-charge states of the Cooper-pair box. We also calculate charge
noise power spectrum generated by quasiparticle capture and
emission processes.

The lifetime of the even-charge state is an exponentially
distributed random variable corresponding to the homogenous
Poisson process. However, the lifetime distribution of the
odd-charge state may deviate from the exponential one. The
deviations come from two sources - the peculiarity  of the
quasiparticle density of states in a superconductor, and the
possibility of quasiparticle energy relaxation via phonon
emission. The odd-charge-state lifetime distribution function
depends on the ratio of the escape rate of the unequilibrated
quasiparticle from the box $\Gamma_{\rm{out}}$ and quasiparticle
energy relaxation rate $1/\tau$.

The conventional Poissonian statistics for both quasiparticle
entrances to and exits from the Cooper-pair box would lead to a
Lorentzian spectral density $S_Q(\omega)$ of CPB charge
fluctuations~\cite{Machlup}. The interplay of tunneling and
relaxation rates in the exit events may result in deviations from
the Lorentzian function.  In the case of slow quasiparticle
thermalization rate compared to the quasiparticle tunneling out
rate $\Gamma_{\rm out}$, the function $S_Q(\omega)$ roughly can be
viewed as a superposition of two Lorentzians. The width of the
first one is controlled by the processes involving quasiparticle
thermalization and activation by phonons, while the width of the
broader one is of the order of the escape rate $\Gamma_{\rm out}$.
\begin{acknowledgments}

The authors thank A.~Andreev, J.~Aumentado, A.~Ferguson,
A.~Kamenev, O.~ Naaman, D.~Prober, and B.~Shklovskii for
stimulating discussions. The authors acknowledge the hospitality
of Max Planck Institute for the Physics of Complex Systems
(Dresden) where part of this work was done. This work is supported
by NSF grants DMR 02-37296,  and DMR 04-39026.
\end{acknowledgments}

\appendix

\section{Power spectrum of charge noise.}

Combining Eqs.~(\ref{correlations}),~(\ref{sol2}), and~(\ref{L})
we obtain the expression for the charge noise power spectrum,
\begin{widetext}
\begin{eqnarray}\label{noise_app1}
S_Q(\omega)\!&=&\!\frac{4e^2}{\mathcal{L}(\omega)\mathcal{L}(-\omega)}\!\\
\!&\!\times\!&\!\left(\!\sum_{k}\frac{(\omega^2\!+\!\frac{1}{\tau^2})\Gamma_{\rm{out}}(E_k)f(E_k\!-\!\delta
E)\bar{\sigma}_{\rm{even}}\!+\!\Gamma_{\rm{out}}(E_k)^2\rho^b_{\rm{odd}}(E_k)\frac{\bar{\sigma}_{\rm{odd}}}{\tau}}{\left(\omega^2\!+\!(\Gamma_{\rm{out}}(E_k)\!+\!1/\tau)^2\right)}
\!+\!\frac{\bar{\sigma}_{\rm{odd}}}{\tau}\left|\sum_k\frac{\Gamma_{\rm{out}}(E_k)\rho^b_{\rm{odd}}(E_k)}{-i\omega+\Gamma_{\rm{out}}(E_k)+\frac{1}{\tau}}\right|^2\right).\nonumber
\end{eqnarray}
Here the product $\mathcal{L}(\omega)\mathcal{L}(-\omega)$ is
given by
\begin{eqnarray}\label{L_app1}
\mathcal{L}(\omega)\mathcal{L}(-\omega)\!&\!=\!&\!\omega^2\left(1-\frac{1}{\tau}\sum_k\frac{\rho^b_{\rm{odd}}(E_k)\Gamma_{\rm{out}}(E_k)}{\omega^2\!+\!\left(\Gamma_{\rm{out}}(E_k)\!+\!1/\tau\right)^2}\!+\!\sum_k\frac{f(E_k\!-\!\delta
E)\Gamma_{\rm{out}}(E_k)^2}{\omega^2\!+\!\left(\Gamma_{\rm{out}}(E_k)\!+\!1/\tau\right)^2}\right)^2\!+\!\nonumber\\
\!&\!+\!&\!\left(\frac{1}{\tau}\sum_k\frac{\rho^b_{\rm{odd}}(E_k)\Gamma_{\rm{out}}(E_k)(\Gamma_{\rm{out}}(E_k)\!+\!1/\tau)}{\omega^2\!+\!\left(\Gamma_{\rm{out}}(E_k)\!+\!1/\tau\right)^2}\!+\!\sum_k\frac{f(E_k\!-\!\delta
E)\Gamma_{\rm{out}}(E_k)(\omega^2\!+\!1/\tau^2\!+\!\Gamma_{\rm{out}}(E_k)/\tau)}{\omega^2\!+\!\left(\Gamma_{\rm{out}}(E_k)\!+\!1/\tau\right)^2}\right)^2.\nonumber\\
\end{eqnarray}
\end{widetext}
Equation~(\ref{noise_app1}) can be simplified in the thermodynamic
limit by introducing functions $Z_1(\omega)$ and $Z_2(\omega)$
\begin{eqnarray}\label{z1_app1}
Z_1(\omega)=\frac{\Gamma_{\rm{out}}}{D}\sum_k\frac{\rho^b_{\rm{odd}}(E_k)\Gamma_{\rm{out}}(E_k)}{\omega^2\!+\!\left(\Gamma_{\rm{out}}(E_k)\!+\!1/\tau\right)^2},
\end{eqnarray}
and
\begin{eqnarray}\label{z2_app1}
Z_2(\omega)=\frac{1}{D}\sum_k\frac{\rho^b_{\rm{odd}}(E_k)\Gamma_{\rm{out}}(E_k)^2}{\omega^2\!+\!\left(\Gamma_{\rm{out}}(E_k)\!+\!1/\tau\right)^2}.
\end{eqnarray}
Here $C$ and $D$ are given by Eqs.~(\ref{CandD}). Substituting
Eqs.~(\ref{L_app1})~-~(\ref{z2_app1}) into Eq.~(\ref{noise_app1})
one can obtain the general expression for $S_Q(\omega)$
\begin{widetext}
\begin{eqnarray}\label{App_main_noise}
S_Q(\omega)=\frac{4e^2}{\Gamma_{\rm{out}}}\frac{\left[\left(\frac{\omega}{\Gamma_{\rm{out}}}\right)^2\!+\!\left(\frac{1}{\tau\Gamma_{\rm{out}}}\right)^2\right]C
Z_1(\omega)\bar{\sigma}_{\rm{even}}\!+\!D
Z_2(\omega)\frac{\bar{\sigma}_{\rm{odd}}}{\tau\Gamma_{\rm{out}}}-D^2\frac{\bar{\sigma}_{\rm{odd}}}{\tau\Gamma_{\rm{out}}}\left[\left(\frac{Z_1(\omega)}{\tau\Gamma_{\rm{out}}}\!+\!Z_2(\omega)\right)^2\!+\!\left(\frac{\omega}{\Gamma_{\rm{out}}}\right)^2Z_2^{\,2}(\omega)\right]}{\left(\frac{\omega}{\Gamma_{\rm{out}}}\right)^2\left[1-\frac{D}{\tau\Gamma_{\rm{out}}}
Z_1(\omega)\!+\!C
Z_2(\omega)\right]^2\!+\!\left[\frac{D+C}{\tau\Gamma_{\rm{out}}}
Z_2(\omega)\!+\!\left(C
\left(\frac{\omega}{\Gamma_{\rm{out}}}\right)^2\!+\!\frac{C\!+\!D}{(\tau\Gamma_{\rm{out}})^2}\right)Z_1(\omega)\right]^2}.
\end{eqnarray}
\end{widetext}

The functions $Z_1(\omega)$ and $Z_2(\omega)$ can be written in
the form of the dimensionless integrals
\begin{eqnarray}\label{integrals_a}
Z_1(\omega)=\frac{\Gamma_{\rm{odd}}}{\nu(\delta
E)}\int_0^{\infty}dz
\frac{e^{-z}\nu(z)\Gamma_{\rm{odd}}(z)}{\omega^2\!+\!\left(\Gamma_{\rm{odd}}(z)+1/\tau\right)^2},
\end{eqnarray}
and
\begin{eqnarray}\label{integrals_b}
Z_2(\omega)=\frac{1}{\nu(\delta E)}\int_0^{\infty}dz
\frac{e^{-z}\nu(z)\Gamma_{\rm{odd}}^2(z)}{\omega^2\!+\!\left(\Gamma_{\rm{odd}}(z)+1/\tau\right)^2}.\end{eqnarray}
The dimensionless variable $z$ here is defined in Eq.~(\ref{z}).
Assuming that at low temperature the main contribution to the
integrals~(\ref{integrals_a}) and (\ref{integrals_b}) comes from
the small $z$ region, $z\ll \delta E/2T$, one can simplify
$Z_1(\omega)$ and $Z_2(\omega)$ using Eq.~(\ref{smallz}) to obtain
\begin{eqnarray}\label{integrals2}
Z_1(\omega)\approx\int_0^{\infty}dz
\frac{e^{-z}\sqrt{z}}{(\omega/\Gamma_{\rm{odd}})^2z\!+\!\left(1\!+\!\sqrt{z}/\tau\Gamma_{\rm{odd}}\right)^2},\nonumber
\end{eqnarray}
and
\begin{eqnarray}
Z_2(\omega)\approx\int_0^{\infty}dz
\frac{e^{-z}}{(\omega/\Gamma_{\rm{odd}})^2z\!+\!\left(1\!+\!\sqrt{z}/\tau\Gamma_{\rm{odd}}\right)^2}.\nonumber
\end{eqnarray}
In the slow relaxation case $\tau\Gamma_{\rm{odd}}\gg 1$ functions
$Z_1(\omega)$ and $Z_2(\omega)$ are approximately given by
Eqs.~(\ref{z1_2}).

Finally, by taking the appropriate limits in
Eq.~(\ref{App_main_noise}) one can recover
Eq.~(\ref{qual_noise_fast}) for ``deep" and
Eqs.~(\ref{noise_slow_1}) and~(\ref{noise_slow1}) for ``shallow"
traps, respectively.


\begin{thebibliography}{99}
\bibitem{Mannik} J. Mannik and J. E. Lukens, Phys. Rev. Lett. {\bf 92}, 057004 (2004).
\bibitem{Aumentado} J. Aumentado, M. W. Keller, J. M. Martinis,  and M. H. Devoret, Phys. Rev. Lett.
{\bf 92}, 066802 (2004).
\bibitem{Guillaume} A. Guillaume, J. F. Schneiderman, P. Delsing, H. M. Bozler, and
P. M. Echternach, Phys. Rev. B {\bf 69}, 132504 (2004).
\bibitem{Schneiderman} J.F. Schneiderman, P. Delsing, G. Johansson, M.D. Shaw, H.M.
Bozler, and P.M. Echternach, unpublished.
\bibitem{Turek} B. A. Turek, K. W. Lehnert, A. Clerk, D. Gunnarsson, K. Bladh, P.
Delsing, and R. J. Schoelkopf,  Phys. Rev. B {\bf 71}, 193304
(2005).
\bibitem{Lu} W. Lu, Z. Ji, L. Pfeiffer, K. W. West and A. J. Rimberg, Nature (London) \textbf{423}, 422
(2003).
\bibitem{Naaman} O. Naaman and J. Aumentado, Phys. Rev. B {\bf 73}, 172504
(2006).
\bibitem{Ferguson}  A. J. Ferguson, N. A. Court, F. E. Hudson, and R. G.
Clark, Phys. Rev. Lett. {\bf 97}, 106603 (2006).
\bibitem{Eiles} T. M. Eiles, and J. M. Martinis,  Phys. Rev. B {\bf 50},
627 (1994).
\bibitem{Matveev} K. A. Matveev, M. Gisselfalt, L. I. Glazman, M. Jonson, and R. I.
Shekhter, Phys. Rev. Lett. {\bf 70}, 2940 (1993).
\bibitem{Lutchyn2}  R. M. Lutchyn, L. I. Glazman, and A. I. Larkin, Phys. Rev. B {\bf 74}, 064515
(2006).
\bibitem{naaman-private-comm} O. Naaman and J. Aumentado, APS March
Meeting 2005 (X16.00002); (private communication).
\bibitem{Machlup} S. Machlup, J. Appl. Phys. {\bf 25}, 341 (1954).
\bibitem{Lutchyn1}  R. Lutchyn, L. Glazman, and A. Larkin, Phys. Rev. B {\bf 72}, 014517
(2005).
\bibitem{footnote} We assume the system is at low temperature $T\ll
T_b^{*}$, where $T_b^{*}$ is a characteristic temperature at which
thermal quasiparticles appear
$T_b^{*}=\frac{\Delta}{\ln(\Delta/\delta_b)}$.
\bibitem{Schrieffer} J.R. Schrieffer, \textit{Theory of Superconductivity}, (Perseus, Oxford 1999).
\bibitem{Tinkham} M. Tinkham, \textit{Introduction to Superconductivity}, (McGraw-Hill, New York,
1996), p. 81.
\bibitem{Kaplan} S.B. Kaplan \textit{et. al.}, Phys. Rev. B {\bf 14}, 4854
(1976).
\bibitem{Clarke} C.C. Chi and J. Clarke, Phys. Rev. B {\bf 19}, 4495
(1979).
\bibitem{Santhanam} P. Santhanam and D. E. Prober, Phys. Rev. B {\bf 29}, 3733
(1984).
\bibitem{Uren} M.J. Uren, M.J. Kirton, and S. Collins, Phys. Rev. B {\bf 37}, 8346
(1988).
\bibitem{Lax_Mengert} M. Lax, and P. Mengert, J. Phys. Chem. Solids, {\bf 14}, 248 (1960)
\bibitem{Prober} C. M. Wilson, L. Frunzio, and D. E. Prober, Phys. Rev. Lett. \textbf{87}, 067004 (2001)
\bibitem{Kogan} Sh. Kogan, \textit{Electronic Noise and Fluctuations in Solids}, (Cambridge University Press, Cambridge, England,
1996), Sh. M. Kogan and A. Ya. Shul'man, Zh. Eksp. Teor. Fiz. {\bf
56} (1969) 862 [Sov. Phys. JETP {\bf 29}, 467 (1969)]
\bibitem{footnote2} Boltzmann-Langevin approach is adequate for calculating noise at low frequencies $\omega \ll
T$ (classical limit). This frequency domain is broad enough to
include the rates of the relevant processes affecting the noise
spectrum, see Eqs.~(\ref{Gammaout0})-(\ref{tau}).
\bibitem{Lax} M. Lax, Rev. Mod. Phys. {\bf 32}, 25 (1960).
\bibitem{Zwanzig} M. Bixon and R. Zwanzig, Phys. Rev. {\bf 187}, 267
(1969).
\bibitem{Stegun}  M. Abramowitz and I. A. Stegun, \textit{Handbook of Mathematical Functions}, (Dover, New York, 1972).
\end{thebibliography}
\end{document}